%% file: main-0.tex
\documentclass[sigconf]{acmart} 
\usepackage{multirow}
\usepackage{longtable, supertabular}
\usepackage{wrapfig}
\usepackage{setspace}
\usepackage{natbib}
\usepackage{float}
\usepackage{dblfloatfix}  
\usepackage{graphicx}
\usepackage{enumitem,kantlipsum}
\usepackage{multicol}
\usepackage{array}
\usepackage{comment}
\usepackage{color}
\usepackage{subfigure} 
\usepackage{booktabs}
\usepackage{placeins}
\usepackage{soul}
\newcolumntype{P}[1]{>{\centering\arraybackslash}p{#1}}

\usepackage{ulem}

\AtBeginDocument{%
  \providecommand\BibTeX{{%
    \normalfont B\kern-0.5em{\scshape i\kern-0.25em b}\kern-0.8em\TeX}}}

\setcopyright{acmcopyright}
\copyrightyear{2023}
\acmYear{2023}
\acmDOI{XXXXXXX.XXXXXXX}

\acmConference[version]{Make sure to enter the correct
  conference title from your rights confirmation email}{for}{arxiv}
\acmBooktitle{arxiv} 
\acmPrice{15.00}
\acmISBN{978-1-4503-XXXX-X/18/06}

\begin{document}

\title{Challenges and Opportunities for the Design of Smart Speakers}

\author{Tao Long}
\email{long@cs.columbia.edu}
\affiliation{%
  \institution{Columbia University}
  \streetaddress{}
  \city{New York}
  \state{New York}
  \country{USA}
  \postcode{10027}
}

\author{Lydia B. Chilton}
\email{chilton@cs.columbia.edu}
\affiliation{%
  \institution{Columbia University}
  \streetaddress{}
  \city{New York}
  \state{New York}
  \country{USA}
  \postcode{10027}
}

\renewcommand{\shortauthors}{Tao Long and Lydia B. Chilton}

\begin{abstract}
Advances in voice technology and voice user interfaces (VUIs) — such as Alexa, Siri, and Google Home — have opened up the potential for many new types of interaction. However, despite the potential of these devices reflected by the growing market and body of VUI research, there is a lingering sense that the technology is still underused. In this paper, we conducted a systematic literature review of 35 papers to identify and synthesize 127 VUI design guidelines into five themes. Additionally, we conducted semi-structured interviews with 15 smart speaker users to understand their use and non-use of the technology. From the interviews, we distill four design challenges that contribute the most to non-use. Based on their (non-)use, we identify four opportunity spaces for designers to explore such as focusing on information support while multitasking (cooking, driving, childcare, etc), incorporating users' mental models for smart speakers, and integrating calm design principles.
\end{abstract}


\maketitle

\input{main-1intro}
\input{main-2related}

\input{main-3study1method}

\input{main-4study1result}

\input{main-5study2method}
\input{main-6study2result}

\input{main-7study2oppo}

\input{main-8miscs}

\bibliographystyle{ACM-Reference-Format}
\bibliography{main-0}

\end{document}

%% file: main-1intro.tex
\section{Introduction}
Advances in voice technology and voice user interfaces (VUIs) such as smart speakers have opened up the potential for many new types of interaction. Devices such as  Amazon Echo (Alexa), Apple HomePod (Siri), and Google Home, 
use voice interaction in the home to do things that desktops and phones are not designed to do. 
Because they are voice-operated, users can interact with smart speakers while their hands are busy (while dressing, or cooking), or when they have limited visual attention (while driving or taking care of children).
Because they are situated in our home environment, they can help us with environment-related tasks like setting timers while cooking, playing music in the shower, turning off the lights before bed, and  ordering a taxi as you walk out the door.
In many ways, these devices have created the potential for the long-standing HCI visions of usable computing by enabling ubiquitous computing and calm design, which aims at designing technology that is ambient, unobtrusive, and smoothly integrated into daily life experiences \cite{weiser_coming_1996, weiser_computer_nodate}.

Despite the potential of these devices, there is a lingering sense among customers that the technology is still underused. Both marketplace review analysis \cite{fruchter_nonuse} and many user experience studies \cite{mavrina_alexa_2022, alexatoy, voit, kim_exploring_2021, cho_once_2019, whatcanihelp, lopatovska_talk_2019} report that smart speaker users infrequently use the device or abandon it. These studies report multiple potential reasons for the non-use: limited use cases, cumbersome setup experiences, and difficulty in discovering new use cases. However, it is unclear how to design smart speakers better such that their potential might be reached.


To understand the landscape of known challenges in designing smart speakers, we conducted a systematic literature review of 35 research papers containing design principles for VUIs. Together, these papers provide 127 design recommendations and evaluation heuristics. After a thematic analysis, we synthesized the guidelines into five \textbf{themes}: 1) enhance basic usability, 2) customize for user contexts, 3) speak users’ language, 4) design simpler interactions, and 5) establish trust. Each theme contains multiple sub-themes with actionable pathways to realize the theme and improve user experience. These design guidelines are truly important and concrete, providing a solid foundation for the design process of future smart speaker devices.

Although these design principles are all valuable, they sketched a large design space and they did not directly address the most pressing problem of device non-use. The actual design process of smart speakers is still difficult since it's still a large underexplored design space with limited successful applications currently populating as well as a gradual increase of non-use. To prioritize the design guidelines and design space to address the non-use through a user-centered lens, we conducted user interviews with 15 smart speaker users centered on these themes to understand the reason for their use and non-use of the technology. Specifically, we asked about their general experiences with their smart speaker device(s), with an emphasis on what was and was not useful about the device, including their daily interaction with the device, degrees of understanding of smart-speaker-specific concepts, and future expectations. Analyzing the interview data using affinity diagramming, we distilled four major \textbf{challenges} that contribute the most to their non-use of the technology: 1) inefficiency of input and output, 2) lack of capability of handling complex tasks, 3) poor discoverability and 4) misleading mental model.

Based on the successful use cases in the interviews, we present four \textbf{opportunities} for designers in building future smart speakers and addressing non-use. They address some of the key design challenges we identified and incorporate principles of calm design.

This paper makes the following contributions:

\begin{itemize}
    \item Users' mental model of smart speakers as a truly intelligent personal assistant is not aligned with the current capabilities, which leads to disappointment and disuse. Technologically, smart speakers are most capable of issuing simple commands, not engaging in prolonged conversations. Applications should aim for these short interactions and communicate a mental model of \textbf{command-driven interaction}.
    \item Users report many mundane uses of smart speakers (playing music, setting timers, asking for the weather). They see them as trivial and not living up to the potential of the technology. However, these \textbf{seemingly trivial interactions are very useful and consistent with calm design principles} of making computers invisible and helping users stay focused.

    \item A promising point in the design space is to develop applications means to \textbf{provide supporting information while the user is engaged in another task}. For example, getting the weather while the user is getting dressed, playing music while they are in the shower, and setting a timer while they are cooking. These are things that smartphones do not currently assist with, and it suits the command-driven capabilities of smart speakers and follows the calm design principle of helping users stay focused on their current task.
    \item An important challenge in the design space is to eliminate the frequent inefficient interactions. These stem from commands not being misunderstood, device 'talking' too much, and poor navigation. 
    These frictions make the interactions require user attention, which violates calm design principles and cause user lots of frustration. Thus even \textbf{small inefficiencies lead to non-use.}  
    \item Discoverability is a huge challenge to users in finding applications and interactions they will like and use. One opportunity is to \textbf{use short social media videos to help users share their experiences in context}.
\end{itemize}

%% file: main-2related.tex
\section{Related Works}
\subsection{History of VUIs and Smart Speakers}

Smart speakers and voice interaction have long been a grand challenge in human-computer interaction. Starting from the 1930s, Bell Lab and IBM have been building systems for generating continuous human speech \cite{dudley_synthetic_1939} and recognizing digits and spoken words and performing arithmetic operations \cite{noauthor_ibm_2003}. Around the 1980s, significant advancements in speech synthesis and recognition were made thanks to the rise of PCs and Windows as the standard operating system as well as the shifts in speech synthesis and recognition systems, which improved computing power and made everyday speech recognition more achievable \cite{pearl_designing_nodate, dasgupta_introduction_2018, kingaby_voice_2022}. In the early 2000s, VUIs were introduced to the public through interactive voice response systems which operated over telephones \cite{pearl_designing_nodate}. These systems could understand human speech and complete simple tasks like booking flight tickets, refilling medical prescriptions, and transferring money \cite{dasgupta_introduction_2018}.  

Being integrated into mobile devices, voice recognition had improved to the point when Apple introduced Siri in 2011, thus enabling hand-free interaction and popularizing the idea of a voice assistant \cite{gupta_carew_2012, allworth_2014}. Steady improvements in machine learning and the availability of significant datasets fuelling them continually improved the quality of speech recognition and interactions with the technology \cite{siri_team_2017, reeves_ml}. Later, along with the increased access to mobile devices and the advances of cloud computing, people started having more experience with voice assistant technologies embedded in their personal devices \cite{accessphone, terzopoulos_satratzemi_2020}. Providing a wider range of affordances, these embedded VUIs enable users with hands-free interaction and convenience, thus becoming popular \cite{bartle_second_2022}. Until this, people have accessed this emerging technology with only their own phone and their voices \cite{dasgupta_introduction_2018}.

The concept of smart home and modern stand-alone home assistants became widely known to the public as Amazon released its first Amazon Echo in 2014, largely changing the landscape \cite{pearl_designing_nodate, mutchler_2018}. Later, many leading technology companies like Google, Apple, Xiaomi, etc. also released their stand-alone smart speakers to control all smart devices in a household setting \cite{companies_release}. Working actively with users in their daily life by automating simple tasks and enabling hands-free interactions, these new stand-alone smart speakers opened up great potential for new interaction between the users and the voice technology. For instance, various applications of smart speakers appear in the needs of senior care and healthcare \cite{brewer_if_2022, alexatoy, yu_shane_schlosser_o’brien_allen_abramson_flynn_2018, qiu_nurse_2021, 10.1145/3531073.3531157}, companionship and accessibility support \cite{ramadan_amazoncom_2021, 87036379079344, mande_deaf_2021, branham_reading_2019, orlofsky2022older}, education and workspace \cite{daley2020alexa, burns2019alexa, schoegler2020use, liew2022alexa, marikyan2022alexa}, entertainment and hospitality \cite{fan2022talk, buhalis2022voice, cai2022customers, ramadan2021amazon, pollmann2020robot}, e-commerce and customer service \cite{dash2022alexa, lim2022alexa, malodia2022can}, etc. As these companies are vying for control of your home as well as the increasing stay-at-home behavior during the pandemic, more people adopt a device-based home assistant \cite{national_public_radio_npr_use_2020}. In 2017, around 46\% of American adults were using voice assistants while 8\% of Americans were using a separate stand-alone device like Alexa or Google Home \cite{olmstead_nearly_nodate}. In 2022, 62\% of American adults are using voice assistants, and 35\% of American adults are using stand-alone smart speaker devices at home \cite{national_public_radio_npr_npr_2022}.

Today, despite the widespread adoption of voice recognition technology as well as the presence of smart speaker devices at home, there still exist gaps in fulfilling the potential for natural voice interaction.

\subsection{The Vision of Ubiquitous Computing and Calm Technology}

\begin{table*}
   \caption{Principles of Calm Technology by Case \cite{case_calm_2015}}

 \label{table:calmdesign}
   \Description[Principles of Calm Technology by Case \cite{case_calm_2015}]{Principles of Calm Technology by Case \cite{case_calm_2015}.}
    \centering
    \begin{tabular}{l}
    \hline
- C1: Technology should require the smallest possible amount of attention.\\
- C2: Technology should inform and create calm.\\
- C3: Technology should make use of the periphery.\\
- C4: Technology should amplify the best of technology and the best of humanity.\\
- C5: Technology can communicate but doesn’t need to speak.\\
- C6: Technology should work even when it fails.\\
- C7: The right amount of technology is the minimum needed to solve the problem.\\
- C8: Technology should respect social norms.\\
\hline       
\end{tabular}
\end{table*}

Mark Weiser’s vision of ubiquitous computing \cite{weiser_computer_nodate} depicted a future with computers and screens of many sizes embedded into all aspects of our homes and lives: these computers serve in concert to help people effortlessly complete household tasks. Weiser believes that in the future homes, offices, campuses, and cities, there will contain hundreds of these tiny computers around us \cite{weiser_computer_nodate, weiserschool}. Although smart speakers were not the explicit focus of the original vision, it is still consistent with the vision of having quick and easy access to information ready at hand.

However, this proliferation of technology comes with the problem of sensory overload downside: users start feeling anxious \cite{flaherty_2022, lukava_morgado_ramirez_barbareschi_2022}, depressed \cite{lee_son_kim_2016}, and other negative impacts on their mental health \cite{scott_valley_simecka_2016}. Specifically, as users receive more information and have complex interactions with the devices around them, they experience an overload of information, and various sensory inputs \cite{shedroff200011, ulijn_strother_fazal_2012, lee_son_kim_2016}. Experiencing multiple sensory cues (visual, audio, vibration, etc.) at the same time \cite{multisensory}, receiving constant notifications \cite{pielot_notification}, and having long-time online social interaction \cite{matthes_too_2020} all overwhelm our ability to process and respond to information. In addition to the overload, studies show that these burdensome technology experiences may lead to user dissatisfaction \cite{tarafdar_tu_ragu_nathan_2010} and non-use of the technology \cite{ manresa_yee_ponsa_varona_perales_2010}.  

Responding to information and sensory overload created by emerging technologies, Mark Weiser and John Seeley Brown pointed out the potential of \textit{calm technology} \cite{weiser_designing_1995, weiser_coming_1996}. The theory of calm technology states technology should help users focus on the things that are important to them rather than creating panic \cite{weiser_designing_1995, weiser_coming_1996}. Amber Case also published a list of actionable guidelines \cite{case_calm_2015} for designers in designing calm technology. As shown in the following list, most of the guidelines shared the themes of “requiring the smallest possible amount of attention,” “informing and creating calm,” and “making use of the periphery of attention.” Thus, \textbf{the calm technology concept reflects the vision of making computing situated and invisible as well as making users feel calm and focused}.


\begin{figure}[H]
\label{teakettle}
  \centering
  \includegraphics[width=0.9\linewidth]{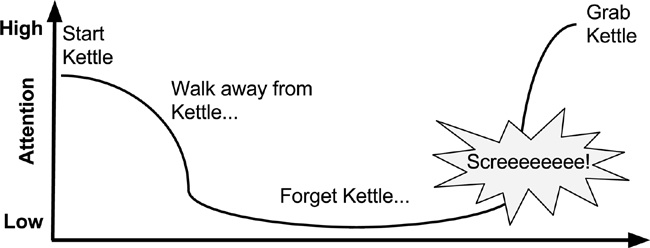}
  \caption{Attention graph for a boiling tea kettle by Case \cite{case_calm_2015}.}
  \Description{The graph shows how much attention is captured by a tea kettle while it's boiling water: the attention is relatively high when the kettle is being set up but diminishes when the user walks away from the kettle, and eventually the kettle is forgotten. When the kettle shouts, all attention is drawn back to the kettle’s state, and the user runs to pick it up}
\end{figure}

An example of the need for calm design principles can be seen in Fig. 1. This attention graph for a tea kettle (See \hyperref[teakettle]{Fig. 1.}) reflects how technology seeks and captures attention from users during the water boiling process: the attention is relatively high when the kettle is being set up but diminishes when the user walks away from the kettle, and eventually the kettle is forgotten. During this period, individuals could redirect their attention to their tasks, like their work, rather than the kettle. Once the water is boiled and the kettle starts to whistle, users' attention is drawn back to it, and they go to pick it up \cite{case_calm_2015}. Therefore, calm design enables users to allows users to prioritize their main tasks over others. The design of a tea kettle, for instance, only requires a small amount of users' attention at the beginning and end. This approach, used in many calm technology designs, suggests that designers should make use of both the center and periphery of our attention to minimize mental load and burden \cite{weiser_designing_1995, weiser_coming_1996, case_calm_2015}. 



\subsection{Mental Models of Smart Speakers}
A key factor in users' interaction with a system -- including it's use or non-use -- comes from their expectations established by a mental model. A mental model refers to the cognitive representation of a system or technology that a user holds in their mind \cite{norman_observations_1983, staggers_mental_1993}.
Users form their mental models unconsciously without prior knowledge or experience with the technology \cite{norman_observations_1983, geromental}. As a result, users must predict how the target system will operate, which shapes their views on the technology's role, purpose, capabilities, and intended uses \cite{geromental, davidoff}. Thus, discrepancies between the mental model and it actually works (the conceptual model \cite{norman_observations_1983, davis_1989} can lead to a gap between the expected and the actual usage of the system, thus causing misunderstanding, misuse, non-use, and abandonment \cite{geromental, manresa_yee_ponsa_varona_perales_2010, nonuserobot, amershi_guidelines_2019}.

The strong personification of these smart speakers — humanized voices and human names — leads to users’ belief that the devices can be modeled as human, thus making users intuitively refer to them as human and assume that it has human intelligence \cite{purington_alexa_2017, mentalmodelemotion}. Additionally, the term “smart” home and the devices' branding as "intelligent assistants" let people have high expectations of the system performance, thus leading to a mismatch between the expectations and the actual capabilities of smart speakers \cite{mavrina_alexa_2022}. New users with little experience with VUIs tend to draw high expectations from their past experiences of human-human interaction \cite{luger_like_2016, cho_once_2019}. However, after finding the system lacks human-like abilities, new users often feel disappointed \cite{alexatoy, whatcanihelp}. Their mental model of the VUIs as having human-level intelligence and capabilities set them up for this disappointment. In conclusion, to solve the problem of non-use, it is crucial to establish a clear understanding of users' mental models and expectations to design effective VUIs that meet users' needs and enhance their experiences.

\noindent
Drawing insights and mapping the future landscape from these previous works, our work seeks to accomplish two main objectives: 1) understand the existing challenges in smart speaker design, especially those that directly cause users to to discontinue use, and 2) identify design opportunities that will enhance the user experience and address issues of non-use in future smart speakers.

%% file: main-3study1method.tex
\begin{figure*}[!b]
  \label{litinfo}
  \centering
  \includegraphics[width=0.99\linewidth]{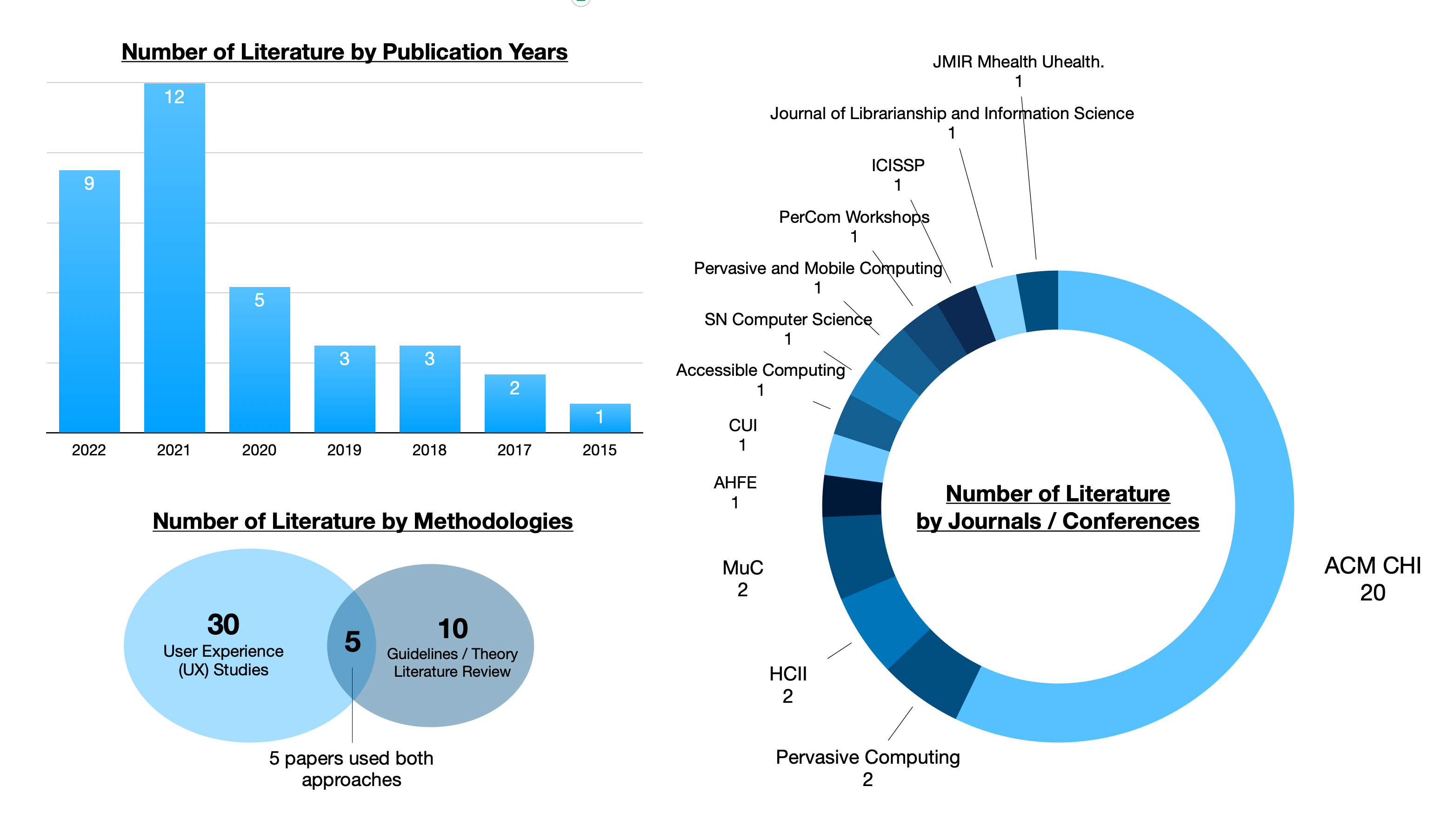}
  \caption{Descriptive data for the 35 papers in the literature review.}
  \Description{Descriptive Data for the 35 Literature. There are three figures: [1]. the image at the top left is a bar plot showing the number of papers published by year: there is a huge surge after 2018. Raw data: 1 paper in 2015, 2 papers in 2017, 3 papers in 2018, 3 papers in 2019, 5 papers in 2020, 12 papers in 2021, and 9 papers in 2022; [2]. the image at the bottom left is a Venn diagram showing the number of literature by research methods. Raw data: 30 papers use user studies and 10 paper use literature reviews (5 papers share both methodologies.); [3]. the right image is a donut/pie chart showing where the studies are published at. Raw data: 19 in ACM CHI, 2 in MuC, 2 in HCII, 2 in Pervasive Computing, and 1 paper in each of the other conference/journals.}
\end{figure*}

\section{Study 1: Literature Review}

\subsection{Goal}
Through a systematic review of 35 related literature, we want to understand what are some shared themes among the existing VUI design and development guidelines. In total, we identified and synthesized 127 guidelines from the papers that either provide design recommendations or evaluation heuristics.

\subsection{Method}
We found 35 related literature that provides future design guidelines for a voice interface, intelligent voice assistant, or Alexa through the ACM Digital Library and Google Scholar.
Firstly, using and combining a variety of terms to form our search queries, “voice interface,” “user experience,” “design,” and “Alexa,” we found an extensive list of 80 full articles. We selected these papers based on their relevance to these terms, and they cover a wide range of human-centered design research topics: designing for marginalized groups, collaboration work, household usage, and personal skill development and training. To map the current landscape of VUI design and to gain insights into specific VUI design guidelines, a second round of paper selection was undertaken. This selection process focused exclusively on papers that contained design guidelines or recommendations and were published after 2014, as this marked the start of the smart speaker era. Following this round, we narrowed it down to 35 full papers that provided clear guidance for the future design of VUIs. 30 out of these 35 papers conducted user studies, through experiments, diary studies, focus groups, semi-structured interviews, etc. 10 of the papers conducted literature reviews to map future design guidelines, and 5 papers shared both methodologies. Also, 21 papers have been published in recent two years.
(See \hyperref[litinfo]{Fig. 2.} for the detailed descriptive data visualization)

All 35 papers provide guidelines: 16 of them generate design recommendations, while 19 of them formulate evaluation heuristics. Finally, by identifying and extracting all the design recommendations and evaluation heuristics, we found 127 VUI design guidelines in total. We put all the guidelines into a shared list and generated a basic codebook by reading through them. We analyzed and synthesized the guidelines to find the key similarities among their directions and visions reflected in their design guidance. Throughout the coding session, we rigorously categorized each guideline into only one category to ensure proper alignment and consistency. The final themes and sub-themes have been iterated over fifteen times to capture best the future vision given by the design guidelines.

%% file: main-4study1result.tex
\subsection{Findings}

From our systematic literature review, we synthesized and generated five major themes that are shared among the 127 design guidelines: 1) enhance basic usability, 2) customize for user contexts, 3) speak users’ language, 4) design for simpler interactions, and 5) establish trust.
Under each theme, there are also sub-themes with a specific approach to realizing the theme. In total, we presented five major themes and fourteen sub-themes. (See \hyperref[littable]{Table 1. Lit Review Themes \& Example Guidelines})
\label{littable}
\input{table_lit.tex}
\newpage
\newpage
\noindent
\textbf{Theme 1. Enhance Basic Usability (51)}
The most frequently-mentioned theme among the guidelines is VUIs' basic usability, which includes user control, error prevention and recovery, discoverability, multimodality, compatibility, etc. These major sub-themes all echoed Nielsen's heuristics \cite{jakob_nielsen_10_1994} and early evaluation heuristics for VUIs \cite{dybkjaer_knowledge_1993, dybkjaer_usability_2001}.

\begin{figure*}[!b]
  \centering
  \includegraphics[width=\linewidth]{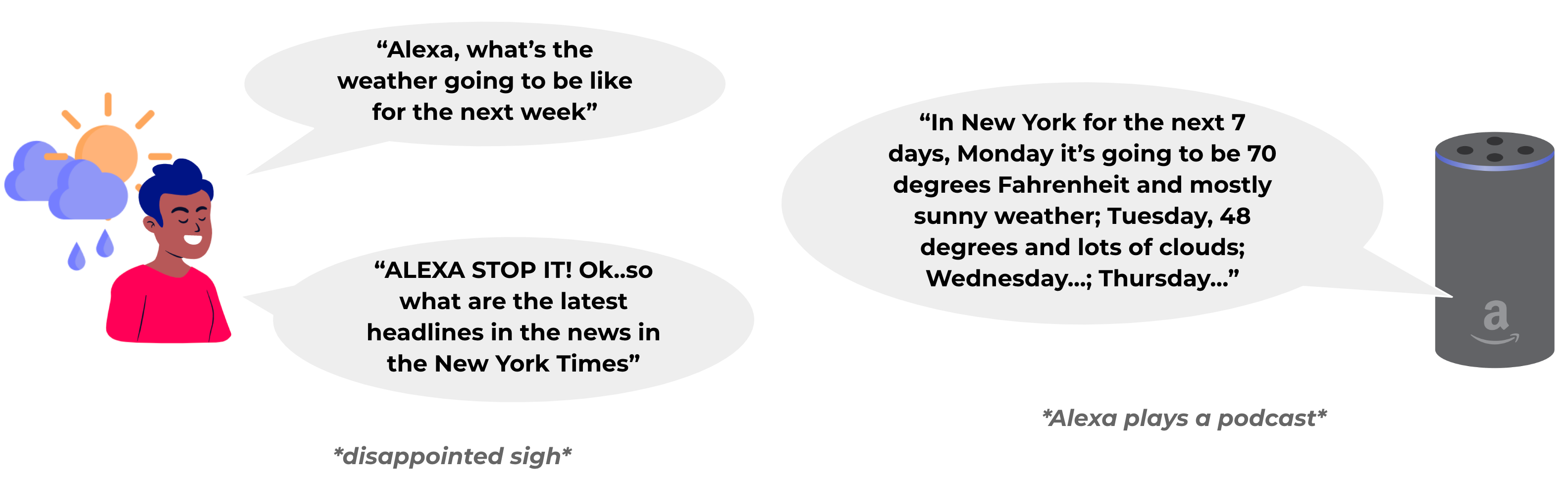}
  \caption{A frustrating interaction with Alexa where the response of the device was first tediously long and second a misunderstanding of the user's command.}
  \Description{A user is having a conversation with Alexa, and they first asked: "Alexa, what’s the weather going to be like for the next week.” Alexa responded: "In New York for the next 7 days, Monday it’s going to be 70 degrees Fahrenheit and mostly sunny weather; Tuesday, 48 degrees and lots of clouds; Wednesday…; Thursday…" Then, the user had to stop it and ask it to play news for them: "ALEXA STOP IT! Ok..so what are the latest headlines in the news in the New York Times." However, Alexa started playing a podcast instead of the news, thus making the user sigh disappointingly.}
  \label{fig:bad_usage}
\end{figure*}

\textbf{(1a)} Promoting error prevention is mentioned in 20 guidelines. Users usually encounter unfamiliarity interacting with VUIs as it is so different from their previous experiences with other technologies \cite{beganymapping}. Thus, designers should help users avoid the potential happening of errors by “making the system status clear \cite{murad_revolution_2019},” “confirming input intelligently \cite{wei_evaluating_2018},” and “handling ambiguous and underspecified utterance \cite{sciutojasonhong}.” Providing VUIs more affordances to prevent errors, users could use the technology more confidently without encountering hardships during their early learning stage, thus prompting their adoption and frequent usage of the VUIs \cite{wei_evaluating_2018}.

 \textbf{(1b)} Strengthening user control over VUIs is mentioned in 13 guidelines. Users could control and operate VUIs according to their preferences and needs: “enable users to share ownership with others \cite{zhang_aware_2022},” “mute or filter notifications \cite{vacher_evaluation_2015},” “provide abilities for users to control and interrupt \cite{maguire_development_2019}.” Offering and establishing a sense of control for users could help reduce errors and ambiguities and is key to generating a sense of security and trust in the system \cite{zhang_aware_2022}.

 \textbf{(1c)} Leveraging the multimodality of VUIs is mentioned in 7 guidelines. Incorporating visual, haptic, and audio interactions could provide users with more information while informing calmness. Visually mark the window that asks for use attention \cite{vacher_evaluation_2015} and “make use of verbal and non-verbal cues \cite{axtell_tea_2021}” provide visual indication and help improve users’ understanding and enrich their experiences with the system \cite{luo_tandemtrack_2020}. 

 \textbf{(1d)} Improving system compatibility is mentioned in 4 guidelines. VUIs should be well integrated and function with users’ frequently-use devices and services \cite{lopez_alexa_2018, sciutojasonhong}, like phones or Apps that people are using daily. Thus, the integration of the services with VUIs largely determines how much money users want to invest and how frequently users will interact with the technology ecosystem \cite{sciutojasonhong}. 

 \textbf{(1e)} Enhancing system discoverability is mentioned in 4 guidelines. Low discoverability means that users cannot easily explore what features, interactions, and experiences are available to them. Due to a lack of a physical screen, VUIs could hardly share a new feature or related information quickly. Thus, leaving much potential by missing helpful interactions, users will just stay with the basic use cases and not learn new things \cite{sciutojasonhong}. Thus, design guidelines mentioned easing the feature-finding process \cite{gollasch_age-related_2021}, utilizing data mining \cite{sciutojasonhong}, using previous responses \cite{striegl}, or giving relevant feature recommendations \cite{kim_exploring_2021} to solve the low discoverability.

 \textbf{(1f)} Facilitating error recovery is mentioned in 3 guidelines. Allowing users to easily “exit from errors or a mistaken conversation \cite{murad_revolution_2019}” helps users avoid getting stuck inside an unsolvable problem or status. It makes users quickly learn the error and keep them in the natural workflow without frustration \cite{setlur_how_2022, goetsu_voice_2019}.

\noindent
\textbf{\\Theme 2. Customize for User Context (26)}
Integrating the usage history, user preferences, and users' background environment helps improve the interaction experiences, thus serving as a huge topic in VUI design guidelines and heuristics. Responding to the mental model of VUIs, users may expect high efficiency and capabilities of communication from them, so designers should incorporate contexts to understand users’ preferences and offer personalized interactions.

\textbf{(2a)} Remembering usage history and applying them to future automation is mentioned in 13 guidelines. Being different from the expectation, users found the VUIs' contextualization is still insufficient as they do not remember many important things as they expect these smart technologies "should be learning". Thus, many guidelines include providing specific accommodations for users' frequently-used commands \cite{commandcus, zhang_aware_2022}. Also, based on users' needs, VUIs should make their interactions more efficient and useful by using context-enabled automation \cite{commandcus, axtell_tea_2021}: "maintaining user profiles to delivering personalized experiences \cite{sabir_hey_2022}" through "associating contents and commands with specific users \cite{zhang_aware_2022}" or "employing territorial markers to help them avoid activity- and preference-related conflicts \cite{zhang_aware_2022}."

\textbf{(2b)} Designing user-centered proactive adaption for special user groups, like parent-kids interactions, seniors, and people with diverse abilities, is mentioned in 9 guidelines. For example, the amount of the VUI automation involved in the interaction \cite{zhang_storybuddy_2022, harrington_its_2022} and responding contents \cite{kim_exploring_2021} should be adapted based on the capabilities and background of the users. Also, designers should utilize the "social-justice oriented design" and spend more effort understanding tools delegated for "the overburdened and under-appreciated workforce \cite{bartle_second_2022}."

\textbf{(2c)} Leveraging the physical background environment is mentioned in 4 guidelines. Similar to Mark Weiser's idea on ubiquitous computing, many guidelines suggest using sensors \cite{reddy_making_2021} and location markers to "modify its level of proactivity, listen for particular commands, and offer more appropriate suggestions for users \cite{zhang_storybuddy_2022}." 

\noindent
\textbf{\\Theme 3. Speak Users’ language (24)}
Many guidelines mentioned that the VUIs should speak users' language including using conversation-like language and expressing human emotions throughout the interaction \cite{yang_understanding_2019}. This topic echoes the users' mental model of smart assistants as they expect high system adaptability to their language. Filling the gap in \cite{gapincalmminimalist}, this theme relates to the consistency and naturalness in Nielsen and early VUI heuristics \cite{jakob_nielsen_10_1994, dybkjaer_usability_2001, suhm_towards_2003}.

\textbf{(3a)} Accommodating conversational language is mentioned in 18 guidelines. Future designers should make VUIs use human-like language and patterns to sound more natural \cite{kim_designers_2021}. These human-like conversation "tricks" includes turn-taking \cite{alrumayh_context_2020, wei_evaluating_2018, murad_finding_2021}, re-mentioning previous contents in the chat history \cite{wang_alexa_2020}, back-channeling \cite{cho_alexa_2022}, and understanding and applying social cues \cite{alrumayh_context_2020}. Human-like conversations and communication strategies will build trust between the users and the system, thus leading to frequent usages \cite{davidoff}.

\textbf{(3b)} Anthropomorphizing human emotions and expressing affective responsiveness are mentioned in 6 guidelines. Designers should make sure VUIs could express interest to users: show empathy and emotional responsiveness \cite{kim_designers_2021, wang_alexa_2020}. As the interaction between users and VUIs becomes more affective and natural, users will become more engaged \cite{kim_designers_2021}.

\noindent
\textbf{\\Theme 4. Design for Simpler Interactions (16)}
Many design recommendations and evaluation heuristics focus on the conciseness and simplicity of the interactions between users and VUIs. Responding to Mark Weiser's vision \cite{weiser_designing_1995, weiser_coming_1996, case_calm_2015}, Amazon's development guidelines, and one-breath test \cite{blankenburg_2018, amazon_alexa_alexa_nodate, amazon_alexa_alexa_nodate1, amazon_alexa_alexa_nodate2}, the conciseness and the simplicity both help users minimize cognitive load \cite{alrumayh_context_2020} and optimize the users' time, thus allowing users to focus on the information that is more crucial \cite{nowacki_improving_2020}. Filling the gap in \cite{gapincalmminimalist}, it also echoes both the "minimalist design" in Nielsen's heuristics \cite{jakob_nielsen_10_1994} and the "the right amount of technology is the minimum needed to solve the problem" in \textit{Calm Technology} \cite{case_calm_2015}.

\textbf{(4a)} Designing for shorter conversations is mentioned in 8 guidelines. Keeping the input and output concise as well as acknowledging users before a long interaction are important \cite{wei_evaluating_2018}. Thus, guidelines mention "maximize efficiency \cite{zwakman_usability_2021}" and "be concise and to the point based on the user’s intent \cite{alrumayh_context_2020}."

\textbf{(4b)} Designing for simple conversation between users and VUIs is mentioned in 8 other guidelines. In addition to time efficiency, VUI designers should keep the interaction structurally \cite{wei_evaluating_2018} and acoustically \cite{zwakman_usability_2021} simple to avoid confusion. Besides, VUIs' responses should also serve guidance \cite{murad_revolution_2019}.

\noindent
\textbf{\\Theme 5. Establish Trust (10)}
As a huge section in security and privacy, establishing trust in VUIs is largely concerned. People are worried that Alexa is always listening and stealing personal data as Alexa is taking advantage of its special setting of homes \cite{lau_alexa_nodate, alwayslisten1, alwayslistening2,  alexareliability}. Establishing a trustworthy perception of VUIs is mentioned in 10 guidelines, from the literature. It includes integrating additional feedback \cite{reddy_making_2021, alrumayh_context_2020}, explicit warning messages and consequences \cite{zubatiy_empowering_2021, chalhoub_it_2021}, tangible controls \cite{kim_designers_2021}, and indicating system contexts between native and third-party skills for users \cite{majordavid_alexa_2021}.


%% file: table_lit.tex



{\renewcommand{\arraystretch}{1.25}

\newpage
{\small
\clearpage
\onecolumn
\begin{longtable}{|P{1.6cm}|P{2.7cm}|p{12.5cm}|}
\captionsetup{justification=centering}
\caption{Lit Review Themes and Example Guidelines
\\ \textit{Note: The number (\#) inside the parentheses indicates how many guidelines belong to the theme.}}\\

\hline
  \textbf{Themes} &
  \textbf{Sub-themes} &
  \textbf{Example Guidelines} \\ \hline
\endfirsthead
\endhead
\multirow{12}{*}{
\begin{tabular}[c]{@{}c@{}}\\\\\\\\\\\\\textbf{Enhance} \\\textbf{Basic}\\ \textbf{Usability}\\ \textbf{(51)}\end{tabular}} &
  \multirow{2}{*}{\begin{tabular}[c]{@{}c@{}}\\Promote\\Error Prevention (20)\end{tabular}} &
 
 A12. Confirm input intelligently: [current VUIs] failed to explicitly confirm some critical actions [, like double-checking which alarm to turn off]  \cite{wei_evaluating_2018}.\\ \cline{3-3} 
 &
   &
   
 [VUIs] need to provide feedback to the user explaining their interpretation of the [ambiguous or underspecified utterance] and how it was handled \cite{setlur_how_2022}.  \\ \cline{2-3} 
 &
  \multirow{2}{*}{\begin{tabular}[c]{@{}c@{}}Strengthen\\ User Control (13)\end{tabular}} &
  Provide ability for users to control and interrupt \cite{vacher_evaluation_2015}. \\ \cline{3-3} 
 &
   &
  The control users have on the processing of their actions by the system. \cite{nowacki_improving_2020}. \\ \cline{2-3} 
 &
  \multirow{2}{*}{\begin{tabular}[c]{@{}c@{}}\\[-0.25cm]Leverage \\Multimodality (7)\end{tabular}} &
  Address {[}low situation awareness{]} by providing better vocal feedback... and adding visual indication for information that is critical to the user \cite{luria} \\ \cline{3-3} 
 &
   &
  Leverage task context and multimodality in order to provide visual or other non-verbal cues \cite{axtell_tea_2021}. \\ \cline{2-3} 
 &
  \multirow{2}{*}{\begin{tabular}[c]{@{}c@{}}\\[-0.25cm]Improve\\ Compatibility (4)\end{tabular}} &
  Smart Home Framework: the compatibility of the [VUIs] with smart home devices \cite{lopez_alexa_2018}. \\ \cline{3-3} 
 &
   &
  Integrate {[}the VUIs with{]} not only smartphones but also connected televisions, computers, and other screen-based devices \cite{sciutojasonhong}. \\ \cline{2-3} 
 &
  \multirow{2}{*}{\begin{tabular}[c]{@{}c@{}}\\Enhance\\ Discoverability (4)\end{tabular}} &
  Use responses to help users discover what is possible...rather than always say something is impossible: the system did not teach ways to ask for a result, and {[}users{]} had to guess and try multiple times \cite{wei_evaluating_2018}. \\ \cline{3-3} 
 &
   &
  Data mining to offer new features: ...significant challenge for [VUIs]. One opportunity is to data mine repeated patterns of use or use common routines as scaffolding to introduce new related features \cite{sciutojasonhong}. \\ \cline{2-3} 
 &
  \multirow{2}{*}{\begin{tabular}[c]{@{}c@{}}\\[-0.25cm]Facilitate\\ Error Recovery (3)\end{tabular}} &
  Provide interface affordances (visual or language) so users can refine and repair system choices \cite{setlur_how_2022}. \\ \cline{3-3} 
 &
   &
  A17. Allow users to exit from errors or a mistaken conversation: Use a special escape word globally (e.g. "Stop")... {[}or{]} non-speech methods when speech fails (e.g., push a physical button) \cite{wei_evaluating_2018}. \\ \hline
\multirow{6}{*}{\begin{tabular}[c]{@{}c@{}}\\\\\\\\\textbf{Customize}\\ \textbf{for User}\\ \textbf{Context}\\ \textbf{(26)}\end{tabular}} &
  \multirow{2}{*}{\begin{tabular}[c]{@{}c@{}}\\ Remember\\Usage History (13)\end{tabular}} &
  Remember User Profiles to Deliver Personalized Services: store a vast amount of information specific to the user, such as personal profiles or preferences \cite{kim_designers_2021} \\ \cline{3-3} 
 &
   &
  Enable user to employ territorial markers to help them avoid activity- and preference-related conflicts by communicating their preferences or staking a claim to the device or data \cite{gargindian}. \\ \cline{2-3} 
 &
  \multirow{2}{*}{\begin{tabular}[c]{@{}c@{}}Design for\\ Diverse \& Sensitive\\ Populations (9)\end{tabular}} &
  Enriching the responding contents when executing these commands might give older adults a chance to find more features and functionalities \cite{kim_exploring_2021}.
  \\ \cline{3-3} 
 &
   &
  Adopt social-justice oriented design methods...when building [VUIs] in home health care contexts \cite{bartle_second_2022}. \\ \cline{2-3} 
 &
  \multirow{2}{*}{\begin{tabular}[c]{@{}c@{}}\\[-0.25cm]Leverage\\ User's\\ Environment (4)\end{tabular}} &
 {Integrate the functionality of ubicomp sensors: install additional sensors in the home that can serve as automatic warning alarms, for example, if the stove is left on or if there is a water leak. {[}or{]} whether someone is within earshot before triggering interactions \cite{reddy_making_2021}}.  \\ \cline{3-3} 
 &
   &
  Leverage knowledge of place: {[}As the VUIs{]} is in a living room versus a bedroom, it can modify its level of proactivity, listen for particular commands, and offer suggestions for new uses \cite{nowacki_improving_2020}. \\ \hline
\multirow{4}{*}{\begin{tabular}[c]{@{}c@{}}\\\\[-0.25cm]\textbf{Speak}\\\textbf{Users'}\\\textbf{Language}\\ \textbf{(24)}\end{tabular}} &
  \multirow{2}{*}{\begin{tabular}[c]{@{}c@{}}\\[-0.25cm] Accommodate\\ Conversational\\ Speech (18)\end{tabular}} &
  Enhancing the message interactivity of the human-{[}VUIs{]} conversation by increasing the degree of contingency in message exchanges \cite{wang_alexa_2020}. \\ \cline{3-3} 
 &
   &
  Tailor responses and follow-up questions to make interactions more engaging, elicit in-depth disclosure, and effectively provide emotional support through these devices most natural responses \cite{shani_alexa_2022}. \\ \cline{2-3} 
 &
  \multirow{2}{*}{\begin{tabular}[c]{@{}c@{}}\\[-0.25cm] Anthropomorphize\\ Human's Emotions (6)\end{tabular}} &
  Employ empathetic expressions to show emotional responsiveness. When [VUIs] use phrases such as "I understand" or "I can relate to you"... users are likely to perceive it being highly social \cite{wang_alexa_2020}. \\ \cline{3-3} 
 &
   &
  Express Sympathy; Be Interesting, Charming, and Lovable; Express Interest to Users \cite{sabir_hey_2022}. \\ \hline
\multirow{4}{*}{\begin{tabular}[c]{@{}c@{}}\\[-0.25cm]\textbf{Design for}\\\textbf{Simpler}\\ \textbf{Interactions}\\ \textbf{(16)}\end{tabular}} &
  \multirow{2}{*}{\begin{tabular}[c]{@{}c@{}}\\[-0.25cm]Design for Shorter\\ Conversation (8)\end{tabular}} &
  \begin{tabular}[c]{@{}l@{}}Design for short interactions, know when it will be long: Systems can prepare for the large majority\\ of interactions to be a single command-answer or command-action \cite{axtell_tea_2021}.\end{tabular} \\ \cline{3-3} 
 &
   &
  A11. Keep feedback and prompts short: ... the {[}current VUIs'{]} responses were not always clear or succinct, making it difficult for users to listen, understand, and remember \cite{wei_evaluating_2018}. \\ \cline{2-3} 
 &
  \multirow{2}{*}{\begin{tabular}[c]{@{}c@{}}Design for Simpler\\ Conversation (8)\end{tabular}} &
Guide users through a conversation so they are not easily lost \cite{murad_revolution_2019}.\\ \cline{3-3} 
 &
   &
    Minimize acoustic confusability of vocabulary \cite{wei_evaluating_2018}.\\ \hline
\multirow{2}{*}{\begin{tabular}[c]{@{}c@{}}\\\textbf{Establish}\\\textbf{Trust}\\ \textbf{(10)}\end{tabular}} &
  \multirow{3}{*}{\begin{tabular}[c]{@{}c@{}}
 \\\\[-0.25cm] Establish\\Trust (10)\end{tabular}} &
 Where available, tangible controls can improve the privacy experience \cite{kim_designers_2021}. \\ \cline{3-3} 
 & &
  Acknowledgments and confirmations: To build trust, acknowledgments need to be provided as feedback indicating that the user’s input was received \cite{alrumayh_context_2020}. \\ \cline{3-3} 
  & &
  Explicitly including a warning message about a third-party vendor in Alexa’s initial response to activating a third-party skill \cite{zubatiy_empowering_2021}. \\

  \hline
\end{longtable}
\twocolumn
}
}

%% file: main-5study2method.tex
\section{Study 2: User Interviews}
\subsection{Goal}
To complement our understanding from the previous systematic literature review on existing guidelines for designing better VUIs, we conducted 15 in-depth interviews with Alexa users. The aim of this study was to better understand the \textbf{challenges} that contribute to the non-use of the technology and identify \textbf{opportunities} for future improvement, with a user-centered and evidence-based lens. Through these interviews, we distilled insights into both the obstacles faced by users and the interactions they enjoyed, allowing us to map future opportunities to design better smart speakers.

\subsection{Method}

Between October 2021 and February 2022, we conducted semi-structured interviews with 15 users of Alexa, who actively use Alexa in their daily life. We specifically chose Alexa to formulate our interview because it now shares the largest market share and it has been the leading brand of VUIs and smart speakers~\cite{reportlinker_smart_2022}. 

We reached out and recruited the participants through a college-wide email network that enabled connections with people from various backgrounds and age groups. Out of the 15 interviews conducted, 4 participants declined to disclose their demographic information. The remaining 11 participants had an average age of 26.1 years, 7 were female, 3 were male, and 1 were gender-variant. On average, they had 4.2 years of experience using Alexa, and 8 of them interacted with the device at least once a day. Each interview session lasted for one hour. With exemption from the Institutional Review Board (IRB), we paid each participant \$20 for their time. We anonymized all the names and identifiable information of the participants. Participants also received a consent form before their interview.  With their consent, the interviews were conducted online through Zoom where we audio recorded and transcribed them. The user interviews aimed to understand the users’ experiences with Alexa, emphasizing what was and was not useful or enjoyable about the device. Specifically, the interview process asked about their daily routine with Alexa, their enjoyment and frustration with Alexa, degrees of understanding of Alexa and Alexa-specific concepts like skills, and future expectations.

After conducting the interview, two individuals coded the interviews and picked the standing-out quotes. Then, they generated codes to find the overarching themes of the challenges and enjoyable moments they encountered during their usage. Finally, they list all the quotes on an affinity diagram that contains all the themes shared among users.

%% file: main-6study2result.tex
\subsection{Findings: Challenges}

From the user study, we found four major challenges that specifically contribute to users’ non-use: 1) inefficiency of input and output, 2) lack of capability for complex tasks, 3) poor discoverability, and 4) misleading mental model. Mapping from these challenges that lead to non-use as well as their enjoyable experiences with smart speakers, we generated a list of design opportunities to improve user experience and solve the abandonment.

\noindent
\textbf{\\Challenge 1: Inefficiency of Input and Output}
\\11 of the 15 users reported frustrating experiences when completing even simple tasks with Alexa. Figure \ref{fig:bad_usage} shows an example. First, the user asked for the weather and got an answer that was tediously long to listen to, so the user yelled at Alexa to stop talking. Then when they asked for news headlines, it misunderstood the intent and played a podcast instead. Misunderstandings and long responses are just two of the types of inefficiency that make interactions frustrating. 
Other inefficiencies shared by the users included: Alexa speaks too slowly (P10), Alexa gave follow-up questions or hints that were not useful and burdensome to them (P10, P11, and P14), and they had to issue the same command multiple times in order to be understood due to accent and culture-specific terms (P6, P7, P12, and P15). 

The problem with input and output inefficiencies was that Alexa forces users to focus on computation rather than the task. Users had to stop Alexa once it had triggered the wrong command. Users had to wait or interrupt their talks for too long. Users had to repeat or re-adjust their commands when they were misunderstood (P3 and P8). These interactions largely required users' energy and attention, which goes against calm design principles (C1). Often, this frustration caused the user to give up on the query if it is not worth the frustration (P9).

The inefficiencies also cost users time and bring negative emotions. Although this time may only be an additional second or two, users reported saying these inefficiencies slowed them down (P5 and P9) and felt frustrated and annoyed (P9, P10, and P14). Nevertheless, it prompts users to explore alternative solutions. For example, P5 shared their experience of trying to use Alexa to play Spotify but ultimately switched back to their trusted method of using their phone due to disappointment with Alexa's inefficiency.
\begin{quotation}
\emph{I'll ask [Alexa] to play a song, and she won't recognize it. I know I'm being impatient...It takes time [for Alexa] to take in what I said... That's why I just [play the music directly] from my phone."} (P5)
\end{quotation}

Thus, echoing the discussion on design for simpler interactions (Literature Review Theme 4), the inefficiency of input and output is a key contributor to non-use. The frustration and annoyance users experienced in unwanted interactions led them to abandon the task or to find easier ways of doing it.

\noindent
\textbf{\\Challenge 2: Lack of Capability for Complex Tasks}
\\Six of the 15 users preferred to use Alexa for simple tasks over complex ones.   
For example, P3 mentioned that they liked using Alexa for the same five or six basic functions, such as turning on lights. In contrast, they found that more complicated requests often led to a longer conversation with multiple back-and-forths with Alexa, which could easily lead to misunderstandings (P1, P7, and P9). Users found it frustrating and upsetting when Alexa could not understand their requests and gave no response (P15) or performed the wrong action (P3 and P14). Besides, when these mistakes happened, the cost was high -- they had to repeat the entire process without any easy way to clarify or correct Alexa:

\begin{quotation}
\emph{“[I] wouldn’t trust her to do a very complex search. She would misunderstand something along the way. She would interpret something else or would get the words wrong... There's no way to easily go back. There’s no “let me go back and clarify.”} (P3)
\end{quotation}

Therefore, the limitations in Alexa's capacity to execute complicated tasks, combined with its lack of robust error prevention and recovery capabilities, led to user frustration and non-use. The effort and time needed by users to restart conversations and stop incorrect actions could lead to disappointment with Alexa's performance, causing them to give up on Alexa and look for alternative solutions to complete their tasks. 

\noindent
\textbf{\\Challenge 3: Poor Discoverability}
\\Nine of the 15 users stated their struggle with Alexa's discoverability: they felt they had only used the most basic functions though they knew that there were many useful Alexa skills existing out there (P1, P2, P3, P5, P9, and P14).
Even though users wanted to explore new skills as well as new use cases, they still found it challenging to locate resources that help them do so:

\begin{quotation}
\emph{“No one really tells, you as a user, what other existing skills there are in Alexa out of the box.”} (P14)
\end{quotation}

Alexa had many avenues to help users discover new skills, but none of them seemed to solve the problem. It suggested skills after a user issues a command, but users disliked this strongly, as it interfered with the efficient execution of their tasks and was often unrelated (See Challenge 1). Alexa devices with screens suggested commands and skills in less obtrusive ways \cite{ambient_2011, ambientdisplay, calmcho, calmjafari}, but this option was not available to all users.
Besides, there is a "skill store" to browse for Alexa skills, but many skills have poor ratings or no ratings. Also, many users didn't even know that Alexa applications were called skills (P4, P5, P10, P12, and P15). Users also shared that they omitted the presence of the Alexa app on their phones and don't understand how it works (P4, P5, and P7). Even if users did know about skills and the skill store, it would require conscious efforts to search for skills.

Discoverability is an issue already identified in the literature as a part of ``enhancing basic usability'' (Literature Review Theme 1). Of the many aspects of basic usability that smart speakers need, discoverability is crucial because if users don't discover the features that will help them, they will under-use the device or abandon it entirely.

\noindent
\textbf{\\Challenge 4: Misleading Mental Model}
\\Five of the 15 users mentioned that they did not think they were using Alexa to its full capacity. They seemed to think it was capable of complex tasks but that they could not access them. For example:
\begin{quotation}
\emph{"I only use it for the most basic functions, [but I know] it can do all these things and I use it for the most basic things.
"} (P2)
\end{quotation}

P1 shared a similar feeling: she knew Alexa was capable of so much more, and she wanted to ``become a better user'' and get to use ``the real capability of Alexa''. Lurking behind these comments was the feeling that the users were only using the "basic" interactions, but there existed more powerful or complex interactions they could be having. Users like P8 had the impression that the device was capable of more sophisticated interactions like being a personal assistant or holding a conversation. It is unclear where users got this expectation. It could be from the branding of it being a "smart" device or an "intelligent assistant." It could be because the device has a human voice, and thus people expect it to have human-level intelligence.  
Either way, users had a mental model of the device as being more capable than it is, and this misleading mental model set them us for disappointment and led them to blame themselves for their failure to find or access the more sophisticated features (P5 and P6).

This misalignment between their mental model of Alexa and its true capabilities might directly lead to non-use since it leaves users dissatisfied with their interaction which over time discourages their usages \cite{manresa_yee_ponsa_varona_perales_2010}.
Thus, helping users have a better mental model more aligned with its capabilities could lead to more user satisfaction and stave off the temptation to abandon it.

%% file: main-7study2oppo.tex
\subsection{Findings: Opportunities}
Responding to the challenge categories, we also develop four related opportunities for future smart speaker designers and developers. Exploring further design directions, we aim to bridge the gap between users’ needs or expectations and the future evolution of the Alexa ecosystem. Specifically, we integrate principles of calm design into future opportunities for smart speakers, as smart speakers' concept of ambient computing aligns well with the vision of calm technology and ubiquitous computing. Along with these design opportunities, we envision future interactions with smart speakers to be efficient and supportive of users' focus and calmness.

\noindent
\textbf{\\Opportunity 1. Explore Use Cases that Provide  Supporting Information while doing a Primary Task}
\\Eight users reported that they found their interactions with Alexa useful when they asked the device to provide supporting information while they performed primary tasks in the physical world.

\begin{figure*}[!hb]
  \centering
  \includegraphics[width=0.99\linewidth]{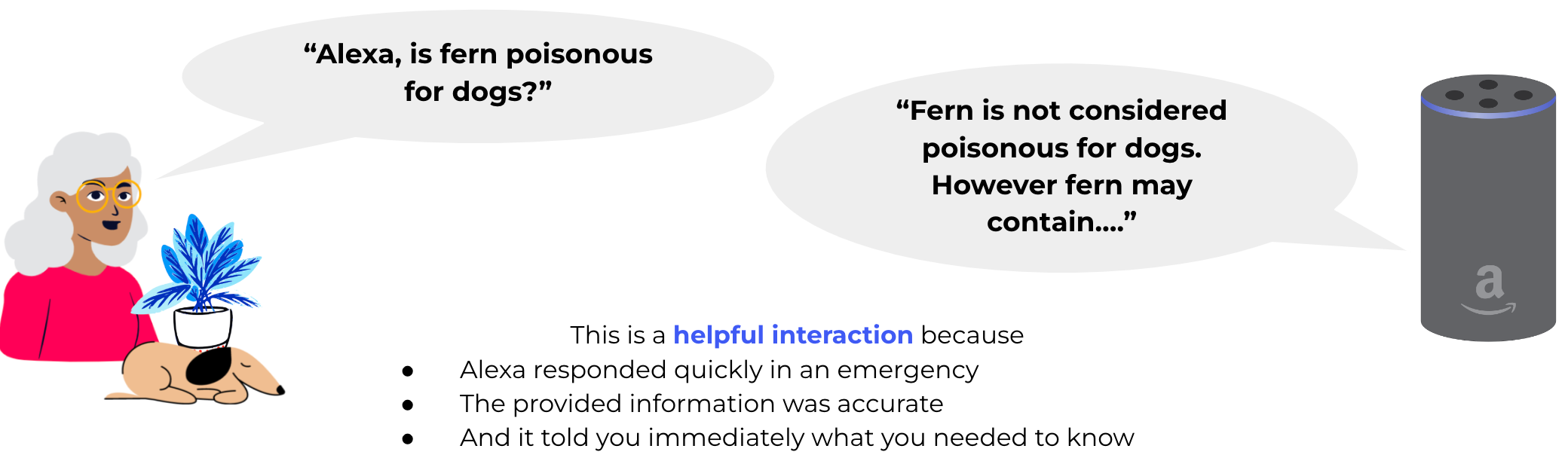}
  \caption{A fulfilling interaction with Alexa where the user needs to find information in an emergency.}
  \Description{The user's dog just ate fern, and the user attended to their dog, while they also wanted to know whether fern is poisonous for their dog. Instead of pulling out their phones, the user asked Alexa: "Alexa is fern poisonous for dogs?" And, Alexa quickly responded, "Fern is not considered poisonous for dogs. However, fern may contain…"}
  \label{fig:dogfern}
\end{figure*}

This included playing music while getting up (P15), asking for the weather while dressing (P12), listening to the news while making breakfast (P3), setting timers while cooking (P13), etc. In most of these tasks, there exists an information need during situational impairments: users' hands or eyes were occupied, making it difficult to complete the task with a phone or a laptop. Opening new ways of interaction by voice, Alexa supports users' needs for information while they are doing a primary task. 
For example, Figure \ref{fig:dogfern} demonstrates the usefulness of smart speakers during emergency situations, such as when a user has to attend to their sick dog and find information about whether fern is poisonous. The traditional method of searching for information on a phone could be time-consuming and require multiple steps, including locating the phone, retrieving it, opening a browser, and using a search engine. Furthermore, the user must remain focused on monitoring the dog, which makes it challenging to perform the task of searching for information. In such situations, smart speakers like Alexa can be of great help in filling the gap by providing quick and easy access to information. In addition, the user can focus on the more important task of caring for their sick dog without having to compromise on obtaining the necessary information. 

Providing information to users doing a primary task fits with the vision of calm design - 
help people stay focused on completing tasks, rather than on computing (C2 and C4). 
By eliminating the need for a phone, it can use the minimum amount of technology needed to solve the problem (C7).
We believe this is a large and unexplored design space as such situations are very common. They can occur in many areas of the house - kitchen bedroom, living room, garage, and even bathroom. Additionally, they can occur to different people doing their tasks - parents, children, older adults, and people with diverse abilities, while all have different needs, tasks, capabilities, and situational impairments - driving, taking care of kids, cooking, emergency situations, even just being in a hurry (late for work, etc). Also, as the challenges of input inefficiency and task complexity pointed out, more targeted and scenario-specific design accommodations could help users get the information they want more efficiently and effortlessly. It echoes the theme of ``design for user context and diverse population'' (Literature Review Theme 2) since this approach helps understand and design for special use cases under different contexts. To accommodate and help users stay focused and receive wanted information, designers should explore how to weave user interaction with smart speakers under these various scenarios. 

Below is a list of possible types of information users may want while doing a primary task:
\label{infowhiletask}
 \begin{itemize}
    \item Living Room
        \begin{itemize}
            \item News \textit{while} having breakfast
            \item Traffic updates \textit{while} having breakfast
            \item Stock updates \textit{while} cleaning the coffee table
            \item Chance of rain \textit{while} heading out of home
            \item Jokes \textit{while} entertaining guests 
            \item Movie recommendations \textit{while} laying on the couch
            \item Home security \textit{while} watching TV
            \item Light control \textit{while} leaving the space
        \end{itemize}
        
    \item Bedroom
         \begin{itemize}
            \item Weather \textit{while} getting dressed
            \item Education and learning resources \textit{while} studying 
            \item Calendar items for the next day \textit{while} being ready for bed
            \item Sport game updates \textit{while} tidying up apartment
            \item Bed stories for kids \textit{while} during bedtime routine 
            \item Guided meditation \textit{while} preparing to sleep
         \end{itemize}
         
    \item Kitchen
         \begin{itemize}
            \item Sports scores \textit{while} washing dishes
            \item Adding items to the shopping list \textit{while} cleaning the fridge
            \item Nutrition information \textit{while} making smoothie
            \item Recipe information \textit{while} cooking
            \item Ingredient substitute \textit{while} baking
         \end{itemize}
         
    \item Restroom
        \begin{itemize}
            \item Music \textit{while} taking a shower
            \item Reminders for appointments \textit{while} touching up makeup
            \item Podcast \textit{while} using the restroom
        \end{itemize}
        
    \item Garden, Garage, and Other Spaces 
         \begin{itemize}
            \item Pet care advice \textit{while} playing with pets
            \item Gardening tips \textit{while} taking care of plants
            \item Timer \textit{while} doing yoga exercises
            \item Metronome \textit{while} practicing guitar
        \end{itemize}
\end{itemize}


\noindent
\textbf{Opportunity 2. Focus on Short Commands rather than Extended Conversations}
\\Three users we interviewed found it difficult to have an extended conversation with Alexa. Instead, six users shared that many successful interactions they had followed a simple interaction pattern: the user issued a short command, and Alexa obeyed (P1, P5, P6, P8, P9, and P14). For example, commands such as ``turn on the light,” ``play Disney Classics on Spotify,” ``what's today's weather,"``set a 10-minute timer for pasta", and ``is ficus dangerous for dogs.” These commands are quick for users to speak, and require short responses with no follow-ups from Alexa. Often the response is a simple noise to acknowledge the command has been processed - like the ding to acknowledge when the lights are turned on. 

These command-driven interactions are consistent with review: designing for simpler interactions (Literature Review Theme 4). 
Moreover, this opportunity consistently aligns with the principles of calm design: technology should demand the smallest possible and right amount of attention (C1 and C7). By having short commands and short Alexa responses, very little attention is required of the user. Also, sometimes Alexa can remain silent (C5), as is the case with Brief Mode \cite{stables_2022}, though none of the users was aware of its existence.
Additionally, Alexa is currently not capable of having long conversations. Thus command-driven interactions help avoid the frustration from Alexa's misunderstandings and difficulty in recovering from errors (Challenge 1).


Although these command-driven interactions are common and useful, users often dismiss them as trivial and not living up to the potential of the technology.
A potential reason for this is users’ mental model of smart speakers as truly intelligent personal assistants regarding their input/output efficiency as well as their capability to handle complex tasks (Challenge 2). This mismatch of expectations is not aligned with the current capabilities, which leads to disappointment and non-use. In conclusion, there may be a need to align Alexa's usage with command-driven interactions, despite it not currently being a part of users' mental model (Challenge 4).

\noindent
\textbf{\\Opportunity 3. Improve Input Efficiency by Learning User Preferences}
\\The efficiency of input (Challenge 1 and Literature Review Theme 4) is one of the biggest problems with Alexa. It often misunderstands users, incorrectly completes a task, has overly lengthy responses, and can't recover from errors well. A potential solution to this is to reduce the need for users to interact with Alexa by learning their preferences and adapting to users' behaviors.
For responses that are overly lengthy, devices should learn over time how to shorten them to provide only the information users need. For example, when the user asks about the weather, Alexa probably doesn't need to say the degrees are "in Fahrenheit" during every daily usage. 
Additionally, if the user often asks for a follow-up about when it is going to rain, that should be added to the weather report, so the user doesn't have to issue the follow-up request.
For commands that Alexa frequently misunderstands, it could use the users' history to make better guesses at what the user wanted based on history. Many people use Alexa as part of their daily routine, and issue the same set of commands daily  such as ``Play workout playlist on Spotify.'' which makes it particularly frustrating when they are misunderstood. Personalization based on usage history has the potential to address this. Such an ambient and passive way of enhancing technology experiences echoes C3 and C4. One case where Alexa already does this is with shopping. Alexa knows the items you've previously purchased, and if you say ``Buy laundry detergent'', it suggests the product you have previously bought rather than having a multi-turn exchange to navigate all the possible detergent options. We believe there is potential for leveraging usage history to make almost every command more efficient.


This opportunity closely relates to the theme of customizing for user contexts and designing shorter and simpler interactions (Literature Review Theme 2 and 4). Also, leveraging usage history and environment to support efficiency, Alexa could directly start accommodating users according to their past experiences and not repeat the same mistakes. This would also help solve the inefficiency of input and output (Challenge 1).
\noindent
\textbf{Opportunity 4. Enhance Discoverability by Promoting Social Learning from Observations and Videos}
\\Discovering applications for Alexa is key to users' usage and satisfaction. Yet discoverability is one of the biggest challenges reported by our users (Challenge 3) and previous studies (Literature Review Theme 1).
A positive example of of discoverability was seeing short videos of people using Alexa in content.
P2 mentions that they adopted Alexa commands from seeing advertisement videos of Alexa: \textit{“there is a guy in the kitchen showing all the things he was doing with his Alexa while cooking. That’s how I learned.”} Also, P10 mentioned seeing videos on video-based social media platforms like TikTok, YouTube, and Instagram, which provides many interesting tricks in using Alexa. For example, TikTok has popular tags such as \#alexa, \#amazonalexa, \#alexahack, \#alexatricks that have received 9.5 billion, 624.7 million, 87.8 million, and 14.6 million views by February 2023. Many viewers of the videos with these tags commented that they had tried the tricks shown and found them both useful and entertaining. 
Similarly, social networks and groups run by smart speaker users like Reddit, StackOverflow, or brand-specific forums like Alexa Developer forums are places for sharing usage experiences, problems, and tricks of smart speakers. 
Thus, despite the challenges of traditional `shoulder-surfing' to  learn applications in informal social situations, video-based social media is an opportunity to share them, and aid discoverability.



%% file: main-8miscs.tex
\section{Limitations and Future Work}
In study 1, we aim to find the common themes among design guidelines for VUIs. Using our methodology of querying for academic papers, we identified 35 papers with guidelines. However, we have undoubtedly missed some. Another source of design guidelines to incorporate would be those made by practitioners, or even by the manufacturers of smart home devices \cite{amazon_alexa_alexa_nodate, amazon_alexa_alexa_nodate1, amazon_alexa_alexa_nodate2, 10.1145/3491102.3517623}. For example, the Alexa developer guide includes the ``one breath test \cite{blankenburg_2018}'' - that everything Alexa says should take only one breadth's worth of words. This limits the length of responses to a duration that users can easily listen to without getting overwhelmed. This is a very handy rule of thumb that was not in any of the design guidelines in academic papers. However, it does fall under the category of Design for Simpler interactions (Theme 4). Whereas the list of guidelines will never be complete, the themes will hopefully be more stable. However, we fully expect them to grow over time.

In the interviews for study 2, we focused on only users of Alexa devices. Thus, it's possible there are challenges and opportunities specific to the smart speakers of other brands. Seven of our users had multiple brand speakers like Google Home and reported similar (but not quite the same) problems (P1, P6, P7, P10, P11, P12, and P13). 

Additionally, not all of our interviewees had the same Alexa device. Notably, some devices have screens while others do not. The Alexa with screens has the potential to aid discoverability by promoting new skills in less obtrusive ways \cite{calmjafari, calmcho, ambient_2011, ambientdisplay, kucera_probing_2017}. Ambient displays use the periphery of our attention, which adheres to the calm design principle 3 (Technology should make use of the periphery). We did not focus on the challenges and opportunities afforded by smart speakers with screens, but this is a promising direction for future work. 

Privacy and trust are known issues with smart speakers and were mentioned in the design guidelines from the literature \cite{jin2022exploring, lau2018alexa, pfeifle2018alexa, sun2020alexa, cho2020will}. However, they aren't a central challenge in our interviews. In part, because the users that already have Alexa have overcome whatever privacy concerns they had already - many people explicitly mentioned being willing to give up privacy to use Alexa (P3, P6, P7, P12, and P15). This is a rich area for research -- particularly for understanding why people don't buy smart speakers, but it's less relevant to understanding their non-use, which is the focus of this work.

Similarly, some other problems shared by participants are not explored in this paper as they do not directly lead to non-use but still deserve further research. For example, eight users encountered many difficulties during the system setup stage because there wasn’t enough information or previous experience in setting up a technology without a visual interface (P1, P3, P5, P6, P9, P10, P13, and P15). Three users also mentioned Alexa's lack of emotional understanding disappoints them as they were frequently talking to the device (P8, P11, and P12).

\section{Conclusion}
To understand the design challenges that lead to non-use and to map future opportunities, this paper conducted a systematic literature review on 127 design guidelines from 35 publications in designing VUIs and user interviews with 15 users. Findings from the literature review present five major themes: 1) enhance basic usability, 2) customize for user contexts, 3) speak users’ language, 4) design for simpler interactions, and 5) establish trust. From the interview, four challenges are identified in contributing to users' non-use: 1) inefficiency of input and output, 2) lack of capability for complex tasks, 3) poor discoverability, and 4) misleading mental model. Lastly, based on the insights and challenges, we propose four future design opportunities for smart speakers to improve user experiences and avoid user non-use, including focusing on information support while multitasking (cooking, driving, showering, childcare, etc), utilizing social media video for social learning, incorporating users' mental models for VUIs, and integrating calm design principles. Our work maps further vision of calm design along with the concept of smart speakers and smart homes.

\section{Acknowledgement}
Thanks to Mareike Carolin Keller, Ray Banke, and Naina Durga Lavakare, who helped conduct the user study and contribute to the early stage of the project.

%% file: main-0.bbl

\begin{thebibliography}{132}


\ifx \showCODEN    \undefined \def \showCODEN     #1{\unskip}     \fi
\ifx \showDOI      \undefined \def \showDOI       #1{#1}\fi
\ifx \showISBNx    \undefined \def \showISBNx     #1{\unskip}     \fi
\ifx \showISBNxiii \undefined \def \showISBNxiii  #1{\unskip}     \fi
\ifx \showISSN     \undefined \def \showISSN      #1{\unskip}     \fi
\ifx \showLCCN     \undefined \def \showLCCN      #1{\unskip}     \fi
\ifx \shownote     \undefined \def \shownote      #1{#1}          \fi
\ifx \showarticletitle \undefined \def \showarticletitle #1{#1}   \fi
\ifx \showURL      \undefined \def \showURL       {\relax}        \fi
\providecommand\bibfield[2]{#2}
\providecommand\bibinfo[2]{#2}
\providecommand\natexlab[1]{#1}
\providecommand\showeprint[2][]{arXiv:#2}

\bibitem[noa(2003)]%
        {noauthor_ibm_2003}
 \bibinfo{year}{2003}\natexlab{}.
\newblock \bibinfo{title}{{IBM} {Archives}: {IBM} {Shoebox}}.
\newblock
\newblock
\urldef\tempurl%
\url{https://www.ibm.com/ibm/history/exhibits/specialprod1/specialprod1_7.html}
\showURL{%
\tempurl}


\bibitem[ama(nda)]%
        {amazon_alexa_alexa_nodate}
 \bibinfo{year}{n.d.}\natexlab{a}.
\newblock \bibinfo{title}{Alexa {Brand} {Guidelines}}.
\newblock
\newblock
\urldef\tempurl%
\url{https://developer.amazon.com/en-US/alexa/branding/alexa-guidelines.html}
\showURL{%
\tempurl}


\bibitem[ama(ndb)]%
        {amazon_alexa_alexa_nodate1}
 \bibinfo{year}{n.d.}\natexlab{b}.
\newblock \bibinfo{title}{Amazon {Alexa} {Voice} {Design} {Guide}}.
\newblock
\newblock
\urldef\tempurl%
\url{https://developer.amazon.com/fr/designing-for-voice/}
\showURL{%
\tempurl}


\bibitem[ama(ndc)]%
        {amazon_alexa_alexa_nodate2}
 \bibinfo{year}{n.d.}\natexlab{c}.
\newblock \bibinfo{title}{Designing {Dialogs} for {Alexa} {Conversations}}.
\newblock
\newblock
\urldef\tempurl%
\url{https://developer.amazon.com/en-US/alexa/alexa-haus/dialogs-for-ac.html}
\showURL{%
\tempurl}


\bibitem[Allworth(2014)]%
        {allworth_2014}
\bibfield{author}{\bibinfo{person}{James Allworth}.}
  \bibinfo{year}{2014}\natexlab{}.
\newblock \bibinfo{title}{Apple's Siri is as revolutionary as the mac}.
\newblock
\newblock
\urldef\tempurl%
\url{https://hbr.org/2011/10/apples-siri-is-as-revolutionar}
\showURL{%
\tempurl}


\bibitem[Alrumayh et~al\mbox{.}(2020)]%
        {alrumayh_context_2020}
\bibfield{author}{\bibinfo{person}{Abrar~S. Alrumayh},
  \bibinfo{person}{Sarah~M. Lehman}, {and} \bibinfo{person}{Chiu~C. Tan}.}
  \bibinfo{year}{2020}\natexlab{}.
\newblock \showarticletitle{Context aware access control for home voice
  assistant in multi-occupant homes}.
\newblock \bibinfo{journal}{\emph{Pervasive and Mobile Computing}}
  \bibinfo{volume}{67} (\bibinfo{date}{Sept.} \bibinfo{year}{2020}),
  \bibinfo{pages}{101196}.
\newblock
\showISSN{1574-1192}
\urldef\tempurl%
\url{https://doi.org/10.1016/j.pmcj.2020.101196}
\showDOI{\tempurl}


\bibitem[Amershi et~al\mbox{.}(2019)]%
        {amershi_guidelines_2019}
\bibfield{author}{\bibinfo{person}{Saleema Amershi}, \bibinfo{person}{Dan
  Weld}, \bibinfo{person}{Mihaela Vorvoreanu}, \bibinfo{person}{Adam Fourney},
  \bibinfo{person}{Besmira Nushi}, \bibinfo{person}{Penny Collisson},
  \bibinfo{person}{Jina Suh}, \bibinfo{person}{Shamsi Iqbal},
  \bibinfo{person}{Paul~N. Bennett}, \bibinfo{person}{Kori Inkpen},
  \bibinfo{person}{Jaime Teevan}, \bibinfo{person}{Ruth Kikin-Gil}, {and}
  \bibinfo{person}{Eric Horvitz}.} \bibinfo{year}{2019}\natexlab{}.
\newblock \showarticletitle{Guidelines for {Human}-{AI} {Interaction}}. In
  \bibinfo{booktitle}{\emph{Proceedings of the 2019 {CHI} {Conference} on
  {Human} {Factors} in {Computing} {Systems}}}. \bibinfo{publisher}{ACM},
  \bibinfo{address}{Glasgow Scotland Uk}, \bibinfo{pages}{1--13}.
\newblock
\showISBNx{978-1-4503-5970-2}
\urldef\tempurl%
\url{https://doi.org/10.1145/3290605.3300233}
\showDOI{\tempurl}


\bibitem[Axtell and Munteanu(2021)]%
        {axtell_tea_2021}
\bibfield{author}{\bibinfo{person}{Benett Axtell} {and} \bibinfo{person}{Cosmin
  Munteanu}.} \bibinfo{year}{2021}\natexlab{}.
\newblock \showarticletitle{Tea, {Earl} {Grey}, {Hot}: {Designing} {Speech}
  {Interactions} from the {Imagined} {Ideal} of {Star} {Trek}}. In
  \bibinfo{booktitle}{\emph{Proceedings of the 2021 {CHI} {Conference} on
  {Human} {Factors} in {Computing} {Systems}}} \emph{(\bibinfo{series}{{CHI}
  '21})}. \bibinfo{publisher}{Association for Computing Machinery},
  \bibinfo{address}{New York, NY, USA}, \bibinfo{pages}{1--14}.
\newblock
\showISBNx{978-1-4503-8096-6}
\urldef\tempurl%
\url{https://doi.org/10.1145/3411764.3445640}
\showDOI{\tempurl}


\bibitem[Bartle et~al\mbox{.}(2022)]%
        {bartle_second_2022}
\bibfield{author}{\bibinfo{person}{Vince Bartle}, \bibinfo{person}{Janice Lyu},
  \bibinfo{person}{Freesoul El~Shabazz-Thompson}, \bibinfo{person}{Yunmin Oh},
  \bibinfo{person}{Angela~Anqi Chen}, \bibinfo{person}{Yu-Jan Chang},
  \bibinfo{person}{Kenneth Holstein}, {and} \bibinfo{person}{Nicola Dell}.}
  \bibinfo{year}{2022}\natexlab{}.
\newblock \showarticletitle{“{A} {Second} {Voice}”: {Investigating}
  {Opportunities} and {Challenges} for {Interactive} {Voice} {Assistants} to
  {Support} {Home} {Health} {Aides}}. In \bibinfo{booktitle}{\emph{Proceedings
  of the 2022 {CHI} {Conference} on {Human} {Factors} in {Computing}
  {Systems}}} \emph{(\bibinfo{series}{{CHI} '22})}.
  \bibinfo{publisher}{Association for Computing Machinery},
  \bibinfo{address}{New York, NY, USA}, \bibinfo{pages}{1--17}.
\newblock
\showISBNx{978-1-4503-9157-3}
\urldef\tempurl%
\url{https://doi.org/10.1145/3491102.3517683}
\showDOI{\tempurl}


\bibitem[Begany et~al\mbox{.}(2015)]%
        {beganymapping}
\bibfield{author}{\bibinfo{person}{Grace Begany}, \bibinfo{person}{Ning Sa},
  {and} \bibinfo{person}{Xiao-Jun Yuan}.} \bibinfo{year}{2015}\natexlab{}.
\newblock \showarticletitle{Factors Affecting User Perception of a Spoken
  Language vs. Textual Search Interface: A Content Analysis}.
\newblock \bibinfo{journal}{\emph{Interacting with Computers}}
  \bibinfo{volume}{28} (\bibinfo{date}{07} \bibinfo{year}{2015}),
  \bibinfo{pages}{iwv029}.
\newblock
\urldef\tempurl%
\url{https://doi.org/10.1093/iwc/iwv029}
\showDOI{\tempurl}


\bibitem[Blankenburg(2018)]%
        {blankenburg_2018}
\bibfield{author}{\bibinfo{person}{Jeff Blankenburg}.}
  \bibinfo{year}{2018}\natexlab{}.
\newblock \bibinfo{title}{Things Every Alexa Skill Should Do: Pass the
  One-Breath Test}.
\newblock
\newblock
\urldef\tempurl%
\url{https://developer.amazon.com/blogs/alexa/post/531ffdd7-acf3-43ca-9831-9c375b08afe0/things-every-alexa-skill-should-do-pass-the-one-breath-test}
\showURL{%
\tempurl}


\bibitem[Branham and Mukkath~Roy(2019)]%
        {branham_reading_2019}
\bibfield{author}{\bibinfo{person}{Stacy~M. Branham} {and}
  \bibinfo{person}{Antony~Rishin Mukkath~Roy}.}
  \bibinfo{year}{2019}\natexlab{}.
\newblock \showarticletitle{Reading {Between} the {Guidelines}: {How}
  {Commercial} {Voice} {Assistant} {Guidelines} {Hinder} {Accessibility} for
  {Blind} {Users}}. In \bibinfo{booktitle}{\emph{The 21st {International} {ACM}
  {SIGACCESS} {Conference} on {Computers} and {Accessibility}}}.
  \bibinfo{publisher}{ACM}, \bibinfo{address}{Pittsburgh PA USA},
  \bibinfo{pages}{446--458}.
\newblock
\showISBNx{978-1-4503-6676-2}
\urldef\tempurl%
\url{https://doi.org/10.1145/3308561.3353797}
\showDOI{\tempurl}


\bibitem[Brewer(2022)]%
        {brewer_if_2022}
\bibfield{author}{\bibinfo{person}{Robin~N. Brewer}.}
  \bibinfo{year}{2022}\natexlab{}.
\newblock \showarticletitle{“{If} {Alexa} knew the state {I} was in, it would
  cry”: {Older} {Adults}’ {Perspectives} of {Voice} {Assistants} for
  {Health}}. In \bibinfo{booktitle}{\emph{Extended {Abstracts} of the 2022
  {CHI} {Conference} on {Human} {Factors} in {Computing} {Systems}}}
  \emph{(\bibinfo{series}{{CHI} {EA} '22})}. \bibinfo{publisher}{Association
  for Computing Machinery}, \bibinfo{address}{New York, NY, USA},
  \bibinfo{pages}{1--8}.
\newblock
\showISBNx{978-1-4503-9156-6}
\urldef\tempurl%
\url{https://doi.org/10.1145/3491101.3519642}
\showDOI{\tempurl}


\bibitem[Buhalis and Moldavska(2022)]%
        {buhalis2022voice}
\bibfield{author}{\bibinfo{person}{Dimitrios Buhalis} {and}
  \bibinfo{person}{Iuliia Moldavska}.} \bibinfo{year}{2022}\natexlab{}.
\newblock \showarticletitle{Voice assistants in hospitality: using artificial
  intelligence for customer service}.
\newblock \bibinfo{journal}{\emph{Journal of Hospitality and Tourism
  Technology}} \bibinfo{volume}{13}, \bibinfo{number}{3}
  (\bibinfo{year}{2022}), \bibinfo{pages}{386--403}.
\newblock


\bibitem[Burns and Igou(2019)]%
        {burns2019alexa}
\bibfield{author}{\bibinfo{person}{Mary~B Burns} {and} \bibinfo{person}{Amy
  Igou}.} \bibinfo{year}{2019}\natexlab{}.
\newblock \showarticletitle{“Alexa, write an audit opinion”: Adopting
  intelligent virtual assistants in accounting workplaces}.
\newblock \bibinfo{journal}{\emph{Journal of Emerging Technologies in
  Accounting}} \bibinfo{volume}{16}, \bibinfo{number}{1}
  (\bibinfo{year}{2019}), \bibinfo{pages}{81--92}.
\newblock


\bibitem[Cai et~al\mbox{.}(2022)]%
        {cai2022customers}
\bibfield{author}{\bibinfo{person}{Ruiying Cai}, \bibinfo{person}{Lisa~Nicole
  Cain}, {and} \bibinfo{person}{Hyeongjin Jeon}.}
  \bibinfo{year}{2022}\natexlab{}.
\newblock \showarticletitle{Customers’ perceptions of hotel AI-enabled voice
  assistants: does brand matter?}
\newblock \bibinfo{journal}{\emph{International Journal of Contemporary
  Hospitality Management}} \bibinfo{number}{ahead-of-print}
  (\bibinfo{year}{2022}).
\newblock


\bibitem[Case(2015)]%
        {case_calm_2015}
\bibfield{author}{\bibinfo{person}{Amber Case}.}
  \bibinfo{year}{2015}\natexlab{}.
\newblock \bibinfo{booktitle}{\emph{Calm {Technology}: {Principles} and
  {Patterns} for {Non}-{Intrusive} {Design}}}.
\newblock \bibinfo{publisher}{"O'Reilly Media, Inc."}.
\newblock
\showISBNx{978-1-4919-2585-0}
\newblock
\shownote{Google-Books-ID: DZ88CwAAQBAJ}.


\bibitem[Chalhoub et~al\mbox{.}(2021)]%
        {chalhoub_it_2021}
\bibfield{author}{\bibinfo{person}{George Chalhoub}, \bibinfo{person}{Martin~J
  Kraemer}, \bibinfo{person}{Norbert Nthala}, {and} \bibinfo{person}{Ivan
  Flechais}.} \bibinfo{year}{2021}\natexlab{}.
\newblock \showarticletitle{“{It} did not give me an option to decline”:
  {A} {Longitudinal} {Analysis} of the {User} {Experience} of {Security} and
  {Privacy} in {Smart} {Home} {Products}}. In
  \bibinfo{booktitle}{\emph{Proceedings of the 2021 {CHI} {Conference} on
  {Human} {Factors} in {Computing} {Systems}}} \emph{(\bibinfo{series}{{CHI}
  '21})}. \bibinfo{publisher}{Association for Computing Machinery},
  \bibinfo{address}{New York, NY, USA}, \bibinfo{pages}{1--16}.
\newblock
\showISBNx{978-1-4503-8096-6}
\urldef\tempurl%
\url{https://doi.org/10.1145/3411764.3445691}
\showDOI{\tempurl}


\bibitem[Cho et~al\mbox{.}(2022)]%
        {cho_alexa_2022}
\bibfield{author}{\bibinfo{person}{Eugene Cho}, \bibinfo{person}{Nasim
  Motalebi}, \bibinfo{person}{S.~Shyam Sundar}, {and} \bibinfo{person}{Saeed
  Abdullah}.} \bibinfo{year}{2022}\natexlab{}.
\newblock \bibinfo{booktitle}{\emph{Alexa as an {Active} {Listener}: {How}
  {Backchanneling} {Can} {Elicit} {Self}-{Disclosure} and {Promote} {User}
  {Experience}}}.
\newblock \bibinfo{type}{{T}echnical {R}eport}.
\newblock
\urldef\tempurl%
\url{https://doi.org/10.1145/3555164}
\showDOI{\tempurl}
\newblock
\shownote{arXiv:2204.10191 [cs]}.


\bibitem[Cho et~al\mbox{.}(2020)]%
        {cho2020will}
\bibfield{author}{\bibinfo{person}{Eugene Cho}, \bibinfo{person}{S~Shyam
  Sundar}, \bibinfo{person}{Saeed Abdullah}, {and} \bibinfo{person}{Nasim
  Motalebi}.} \bibinfo{year}{2020}\natexlab{}.
\newblock \showarticletitle{Will deleting history make alexa more trustworthy?
  effects of privacy and content customization on user experience of smart
  speakers}. In \bibinfo{booktitle}{\emph{Proceedings of the 2020 CHI
  Conference on Human Factors in Computing Systems}}. \bibinfo{pages}{1--13}.
\newblock


\bibitem[Cho et~al\mbox{.}(2019)]%
        {cho_once_2019}
\bibfield{author}{\bibinfo{person}{Minji Cho}, \bibinfo{person}{Sang-su Lee},
  {and} \bibinfo{person}{Kun-Pyo Lee}.} \bibinfo{year}{2019}\natexlab{}.
\newblock \showarticletitle{Once a {Kind} {Friend} is {Now} a {Thing}:
  {Understanding} {How} {Conversational} {Agents} at {Home} are {Forgotten}}.
  In \bibinfo{booktitle}{\emph{Proceedings of the 2019 on {Designing}
  {Interactive} {Systems} {Conference}}} \emph{(\bibinfo{series}{{DIS} '19})}.
  \bibinfo{publisher}{Association for Computing Machinery},
  \bibinfo{address}{New York, NY, USA}, \bibinfo{pages}{1557--1569}.
\newblock
\showISBNx{978-1-4503-5850-7}
\urldef\tempurl%
\url{https://doi.org/10.1145/3322276.3322332}
\showDOI{\tempurl}


\bibitem[Cho and Saakes(2017)]%
        {calmcho}
\bibfield{author}{\bibinfo{person}{Minjoo Cho} {and} \bibinfo{person}{Daniel
  Saakes}.} \bibinfo{year}{2017}\natexlab{}.
\newblock \showarticletitle{Calm Automaton: A DIY Toolkit for Ambient
  Displays}. In \bibinfo{booktitle}{\emph{Proceedings of the 2017 CHI
  Conference Extended Abstracts on Human Factors in Computing Systems}}
  (Denver, Colorado, USA) \emph{(\bibinfo{series}{CHI EA '17})}.
  \bibinfo{publisher}{Association for Computing Machinery},
  \bibinfo{address}{New York, NY, USA}, \bibinfo{pages}{393–396}.
\newblock
\showISBNx{9781450346566}
\urldef\tempurl%
\url{https://doi.org/10.1145/3027063.3052968}
\showDOI{\tempurl}


\bibitem[Cowan et~al\mbox{.}(2017)]%
        {whatcanihelp}
\bibfield{author}{\bibinfo{person}{Benjamin~R. Cowan}, \bibinfo{person}{Nadia
  Pantidi}, \bibinfo{person}{David Coyle}, \bibinfo{person}{Kellie Morrissey},
  \bibinfo{person}{Peter Clarke}, \bibinfo{person}{Sara Al-Shehri},
  \bibinfo{person}{David Earley}, {and} \bibinfo{person}{Natasha Bandeira}.}
  \bibinfo{year}{2017}\natexlab{}.
\newblock \showarticletitle{"What Can i Help You with?": Infrequent Users'
  Experiences of Intelligent Personal Assistants}. In
  \bibinfo{booktitle}{\emph{Proceedings of the 19th International Conference on
  Human-Computer Interaction with Mobile Devices and Services}} (Vienna,
  Austria) \emph{(\bibinfo{series}{MobileHCI '17})}.
  \bibinfo{publisher}{Association for Computing Machinery},
  \bibinfo{address}{New York, NY, USA}, Article \bibinfo{articleno}{43},
  \bibinfo{numpages}{12}~pages.
\newblock
\showISBNx{9781450350754}
\urldef\tempurl%
\url{https://doi.org/10.1145/3098279.3098539}
\showDOI{\tempurl}


\bibitem[Daley and Pennington(2020)]%
        {daley2020alexa}
\bibfield{author}{\bibinfo{person}{Shawn Daley} {and} \bibinfo{person}{Jon
  Pennington}.} \bibinfo{year}{2020}\natexlab{}.
\newblock \showarticletitle{Alexa the Teacher's Pet? A Review of Research on
  Virtual Assistants in Education}.
\newblock \bibinfo{journal}{\emph{EdMedia+ innovate learning}}
  (\bibinfo{year}{2020}), \bibinfo{pages}{138--146}.
\newblock


\bibitem[Dasgupta(2018)]%
        {dasgupta_introduction_2018}
\bibfield{author}{\bibinfo{person}{Ritwik Dasgupta}.}
  \bibinfo{year}{2018}\natexlab{}.
\newblock \showarticletitle{Introduction to {VUI}}.
\newblock In \bibinfo{booktitle}{\emph{Voice {User} {Interface} {Design}:
  {Moving} from {GUI} to {Mixed} {Modal} {Interaction}}},
  \bibfield{editor}{\bibinfo{person}{Ritwik Dasgupta}} (Ed.).
  \bibinfo{publisher}{Apress}, \bibinfo{address}{Berkeley, CA},
  \bibinfo{pages}{1--11}.
\newblock
\showISBNx{978-1-4842-4125-7}
\urldef\tempurl%
\url{https://doi.org/10.1007/978-1-4842-4125-7_1}
\showDOI{\tempurl}


\bibitem[Dash et~al\mbox{.}(2022)]%
        {dash2022alexa}
\bibfield{author}{\bibinfo{person}{Abhisek Dash}, \bibinfo{person}{Abhijnan
  Chakraborty}, \bibinfo{person}{Saptarshi Ghosh}, \bibinfo{person}{Animesh
  Mukherjee}, {and} \bibinfo{person}{Krishna~P Gummadi}.}
  \bibinfo{year}{2022}\natexlab{}.
\newblock \showarticletitle{Alexa, in you, I trust! Fairness and
  Interpretability Issues in E-commerce Search through Smart Speakers}. In
  \bibinfo{booktitle}{\emph{Proceedings of the ACM Web Conference 2022}}.
  \bibinfo{pages}{3695--3705}.
\newblock


\bibitem[Davidoff et~al\mbox{.}(2006)]%
        {davidoff}
\bibfield{author}{\bibinfo{person}{Scott Davidoff}, \bibinfo{person}{Min~Kyung
  Lee}, \bibinfo{person}{Charles Yiu}, \bibinfo{person}{John Zimmerman}, {and}
  \bibinfo{person}{Anind Dey}.} \bibinfo{year}{2006}\natexlab{}.
\newblock \showarticletitle{Principles of Smart Home Control}.
\newblock \bibinfo{journal}{\emph{UbiComp 2006: Ubiquitous Computing}}
  \bibinfo{volume}{4206}, \bibinfo{pages}{19--34}.
\newblock
\showISBNx{978-3-540-39634-5}
\urldef\tempurl%
\url{https://doi.org/10.1007/11853565_2}
\showDOI{\tempurl}


\bibitem[Davis(1989)]%
        {davis_1989}
\bibfield{author}{\bibinfo{person}{Fred~D. Davis}.}
  \bibinfo{year}{1989}\natexlab{}.
\newblock \showarticletitle{Perceived usefulness, perceived ease of use, and
  user acceptance of Information Technology}.
\newblock \bibinfo{journal}{\emph{MIS Quarterly}} \bibinfo{volume}{13},
  \bibinfo{number}{3} (\bibinfo{year}{1989}), \bibinfo{pages}{319}.
\newblock
\urldef\tempurl%
\url{https://doi.org/10.2307/249008}
\showDOI{\tempurl}


\bibitem[de~Graaf et~al\mbox{.}(2017)]%
        {nonuserobot}
\bibfield{author}{\bibinfo{person}{Maartje de Graaf}, \bibinfo{person}{Somaya
  Ben~Allouch}, {and} \bibinfo{person}{Jan van Dijk}.}
  \bibinfo{year}{2017}\natexlab{}.
\newblock \showarticletitle{Why Do They Refuse to Use My Robot? Reasons for
  Non-Use Derived from a Long-Term Home Study}. In
  \bibinfo{booktitle}{\emph{Proceedings of the 2017 ACM/IEEE International
  Conference on Human-Robot Interaction}} (Vienna, Austria)
  \emph{(\bibinfo{series}{HRI '17})}. \bibinfo{publisher}{Association for
  Computing Machinery}, \bibinfo{address}{New York, NY, USA},
  \bibinfo{pages}{224–233}.
\newblock
\showISBNx{9781450343367}
\urldef\tempurl%
\url{https://doi.org/10.1145/2909824.3020236}
\showDOI{\tempurl}


\bibitem[Dudley et~al\mbox{.}(1939)]%
        {dudley_synthetic_1939}
\bibfield{author}{\bibinfo{person}{Homer Dudley}, \bibinfo{person}{R.~R.
  Riesz}, {and} \bibinfo{person}{S.~S.~A. Watkins}.}
  \bibinfo{year}{1939}\natexlab{}.
\newblock \showarticletitle{A synthetic speaker}.
\newblock \bibinfo{journal}{\emph{Journal of the Franklin Institute}}
  \bibinfo{volume}{227}, \bibinfo{number}{6} (\bibinfo{date}{June}
  \bibinfo{year}{1939}), \bibinfo{pages}{739--764}.
\newblock
\showISSN{0016-0032}
\urldef\tempurl%
\url{https://doi.org/10.1016/S0016-0032(39)90816-1}
\showDOI{\tempurl}


\bibitem[Dybkjær and Bernsen(2001)]%
        {dybkjaer_usability_2001}
\bibfield{author}{\bibinfo{person}{Laila Dybkjær} {and}
  \bibinfo{person}{Niels~Ole Bernsen}.} \bibinfo{year}{2001}\natexlab{}.
\newblock \showarticletitle{Usability evaluation in spoken language dialogue
  systems}. In \bibinfo{booktitle}{\emph{Proceedings of the workshop on
  {Evaluation} for {Language} and {Dialogue} {Systems} - {Volume} 9}}
  \emph{(\bibinfo{series}{{ELDS} '01})}. \bibinfo{publisher}{Association for
  Computational Linguistics}, \bibinfo{address}{USA}, \bibinfo{pages}{1--10}.
\newblock
\urldef\tempurl%
\url{https://doi.org/10.3115/1118053.1118055}
\showDOI{\tempurl}


\bibitem[Dybkjær et~al\mbox{.}(1993)]%
        {dybkjaer_knowledge_1993}
\bibfield{author}{\bibinfo{person}{Laila Dybkjær}, \bibinfo{person}{Niels~Ole
  Bernsen}, {and} \bibinfo{person}{Hans Dybkjær}.}
  \bibinfo{year}{1993}\natexlab{}.
\newblock \showarticletitle{Knowledge acquisition for a constrained speech
  system using {WoZ}}. In \bibinfo{booktitle}{\emph{Sixth {Conference} of the
  {European} {Chapter} of the {Association} for {Computational}
  {Linguistics}}}. \bibinfo{publisher}{Association for Computational
  Linguistics}, \bibinfo{address}{Utrecht, The Netherlands}.
\newblock
\urldef\tempurl%
\url{https://aclanthology.org/E93-1061}
\showURL{%
\tempurl}


\bibitem[Fan et~al\mbox{.}(2022)]%
        {fan2022talk}
\bibfield{author}{\bibinfo{person}{Alei Fan}, \bibinfo{person}{Zhi Lu}, {and}
  \bibinfo{person}{Zhenxing~Eddie Mao}.} \bibinfo{year}{2022}\natexlab{}.
\newblock \showarticletitle{To talk or to touch: Unraveling consumer responses
  to two types of hotel in-room technology}.
\newblock \bibinfo{journal}{\emph{International Journal of Hospitality
  Management}}  \bibinfo{volume}{101} (\bibinfo{year}{2022}),
  \bibinfo{pages}{103112}.
\newblock


\bibitem[Feng et~al\mbox{.}(2016)]%
        {multisensory}
\bibfield{author}{\bibinfo{person}{Mi Feng}, \bibinfo{person}{Arindam Dey},
  {and} \bibinfo{person}{Robert~W. Lindeman}.} \bibinfo{year}{2016}\natexlab{}.
\newblock \showarticletitle{An initial exploration of a multi-sensory design
  space: Tactile support for walking in immersive virtual environments}. In
  \bibinfo{booktitle}{\emph{2016 IEEE Symposium on 3D User Interfaces (3DUI)}}.
  \bibinfo{pages}{95--104}.
\newblock
\urldef\tempurl%
\url{https://doi.org/10.1109/3DUI.2016.7460037}
\showDOI{\tempurl}


\bibitem[Flaherty(2022)]%
        {flaherty_2022}
\bibfield{author}{\bibinfo{person}{Aishling Flaherty}.}
  \bibinfo{year}{2022}\natexlab{}.
\newblock \showarticletitle{The Chemistry Teaching Laboratory: A sensory
  overload vortex for students and instructors?}
\newblock \bibinfo{journal}{\emph{Journal of Chemical Education}}
  \bibinfo{volume}{99}, \bibinfo{number}{4} (\bibinfo{year}{2022}),
  \bibinfo{pages}{1775–1777}.
\newblock
\urldef\tempurl%
\url{https://doi.org/10.1021/acs.jchemed.2c00032}
\showDOI{\tempurl}


\bibitem[Fruchter and Liccardi(2018)]%
        {fruchter_nonuse}
\bibfield{author}{\bibinfo{person}{Nathaniel Fruchter} {and}
  \bibinfo{person}{Ilaria Liccardi}.} \bibinfo{year}{2018}\natexlab{}.
\newblock \showarticletitle{Consumer Attitudes Towards Privacy and Security in
  Home Assistants}. In \bibinfo{booktitle}{\emph{Extended Abstracts of the 2018
  CHI Conference on Human Factors in Computing Systems}} (Montreal QC, Canada)
  \emph{(\bibinfo{series}{CHI EA '18})}. \bibinfo{publisher}{Association for
  Computing Machinery}, \bibinfo{address}{New York, NY, USA},
  \bibinfo{pages}{1–6}.
\newblock
\showISBNx{9781450356213}
\urldef\tempurl%
\url{https://doi.org/10.1145/3170427.3188448}
\showDOI{\tempurl}


\bibitem[Garg(2022)]%
        {gargindian}
\bibfield{author}{\bibinfo{person}{Radhika Garg}.}
  \bibinfo{year}{2022}\natexlab{}.
\newblock \showarticletitle{Supporting the Design of Smart Speakers to Foster a
  Sense of Ownership in Asian Indian Families}. In
  \bibinfo{booktitle}{\emph{Proceedings of the 2022 CHI Conference on Human
  Factors in Computing Systems}} (New Orleans, LA, USA)
  \emph{(\bibinfo{series}{CHI '22})}. \bibinfo{publisher}{Association for
  Computing Machinery}, \bibinfo{address}{New York, NY, USA}, Article
  \bibinfo{articleno}{167}, \bibinfo{numpages}{15}~pages.
\newblock
\showISBNx{9781450391573}
\urldef\tempurl%
\url{https://doi.org/10.1145/3491102.3517680}
\showDOI{\tempurl}


\bibitem[Gero et~al\mbox{.}(2020)]%
        {geromental}
\bibfield{author}{\bibinfo{person}{Katy~Ilonka Gero}, \bibinfo{person}{Zahra
  Ashktorab}, \bibinfo{person}{Casey Dugan}, \bibinfo{person}{Qian Pan},
  \bibinfo{person}{James Johnson}, \bibinfo{person}{Werner Geyer},
  \bibinfo{person}{Maria Ruiz}, \bibinfo{person}{Sarah Miller},
  \bibinfo{person}{David~R. Millen}, \bibinfo{person}{Murray Campbell},
  \bibinfo{person}{Sadhana Kumaravel}, {and} \bibinfo{person}{Wei Zhang}.}
  \bibinfo{year}{2020}\natexlab{}.
\newblock \showarticletitle{Mental Models of AI Agents in a Cooperative Game
  Setting}. In \bibinfo{booktitle}{\emph{Proceedings of the 2020 CHI Conference
  on Human Factors in Computing Systems}} (Honolulu, HI, USA)
  \emph{(\bibinfo{series}{CHI '20})}. \bibinfo{publisher}{Association for
  Computing Machinery}, \bibinfo{address}{New York, NY, USA},
  \bibinfo{pages}{1–12}.
\newblock
\showISBNx{9781450367080}
\urldef\tempurl%
\url{https://doi.org/10.1145/3313831.3376316}
\showDOI{\tempurl}


\bibitem[Goetsu and Sakai(2019)]%
        {goetsu_voice_2019}
\bibfield{author}{\bibinfo{person}{Shiyoh Goetsu} {and}
  \bibinfo{person}{Tetsuya Sakai}.} \bibinfo{year}{2019}\natexlab{}.
\newblock \showarticletitle{Voice {Input} {Interface} {Failures} and
  {Frustration}: {Developer} and {User} {Perspectives}}. In
  \bibinfo{booktitle}{\emph{The {Adjunct} {Publication} of the 32nd {Annual}
  {ACM} {Symposium} on {User} {Interface} {Software} and {Technology}}}.
  \bibinfo{publisher}{ACM}, \bibinfo{address}{New Orleans LA USA},
  \bibinfo{pages}{24--26}.
\newblock
\showISBNx{978-1-4503-6817-9}
\urldef\tempurl%
\url{https://doi.org/10.1145/3332167.3357103}
\showDOI{\tempurl}


\bibitem[Gollasch and Weber(2021)]%
        {gollasch_age-related_2021}
\bibfield{author}{\bibinfo{person}{David Gollasch} {and}
  \bibinfo{person}{Gerhard Weber}.} \bibinfo{year}{2021}\natexlab{}.
\newblock \showarticletitle{Age-{Related} {Differences} in {Preferences} for
  {Using} {Voice} {Assistants}}. In \bibinfo{booktitle}{\emph{Mensch und
  {Computer} 2021}}. \bibinfo{publisher}{ACM}, \bibinfo{address}{Ingolstadt
  Germany}, \bibinfo{pages}{156--167}.
\newblock
\showISBNx{978-1-4503-8645-6}
\urldef\tempurl%
\url{https://doi.org/10.1145/3473856.3473889}
\showDOI{\tempurl}


\bibitem[Gupta and Carew(2012)]%
        {gupta_carew_2012}
\bibfield{author}{\bibinfo{person}{Poornima Gupta} {and}
  \bibinfo{person}{Sinead Carew}.} \bibinfo{year}{2012}\natexlab{}.
\newblock \bibinfo{title}{Apple's Siri puts voice-enabled search in spotlight}.
\newblock
\newblock
\urldef\tempurl%
\url{https://www.reuters.com/article/us-ces-siri/apples-siri-puts-voice-enabled-search-in-spotlight-idUSTRE8081VB20120109}
\showURL{%
\tempurl}


\bibitem[Harrington et~al\mbox{.}(2022)]%
        {harrington_its_2022}
\bibfield{author}{\bibinfo{person}{Christina~N. Harrington},
  \bibinfo{person}{Radhika Garg}, \bibinfo{person}{Amanda Woodward}, {and}
  \bibinfo{person}{Dimitri Williams}.} \bibinfo{year}{2022}\natexlab{}.
\newblock \showarticletitle{“{It}’s {Kind} of {Like} {Code}-{Switching}”:
  {Black} {Older} {Adults}’ {Experiences} with a {Voice} {Assistant} for
  {Health} {Information} {Seeking}}. In \bibinfo{booktitle}{\emph{Proceedings
  of the 2022 {CHI} {Conference} on {Human} {Factors} in {Computing}
  {Systems}}} \emph{(\bibinfo{series}{{CHI} '22})}.
  \bibinfo{publisher}{Association for Computing Machinery},
  \bibinfo{address}{New York, NY, USA}, \bibinfo{pages}{1--15}.
\newblock
\showISBNx{978-1-4503-9157-3}
\urldef\tempurl%
\url{https://doi.org/10.1145/3491102.3501995}
\showDOI{\tempurl}


\bibitem[Isyanto et~al\mbox{.}(2020)]%
        {87036379079344}
\bibfield{author}{\bibinfo{person}{Haris Isyanto}, \bibinfo{person}{Ajib~Setyo
  Arifin}, {and} \bibinfo{person}{Muhammad Suryanegara}.}
  \bibinfo{year}{2020}\natexlab{}.
\newblock \showarticletitle{Design and Implementation of IoT-Based Smart Home
  Voice Commands for disabled people using Google Assistant}. In
  \bibinfo{booktitle}{\emph{2020 International Conference on Smart Technology
  and Applications (ICoSTA)}}. \bibinfo{pages}{1--6}.
\newblock
\urldef\tempurl%
\url{https://doi.org/10.1109/ICoSTA48221.2020.1570613925}
\showDOI{\tempurl}


\bibitem[Jafarinaimi et~al\mbox{.}(2005)]%
        {calmjafari}
\bibfield{author}{\bibinfo{person}{Nassim Jafarinaimi}, \bibinfo{person}{Jodi
  Forlizzi}, \bibinfo{person}{Amy Hurst}, {and} \bibinfo{person}{John
  Zimmerman}.} \bibinfo{year}{2005}\natexlab{}.
\newblock \showarticletitle{Breakaway: An Ambient Display Designed to Change
  Human Behavior}. In \bibinfo{booktitle}{\emph{CHI '05 Extended Abstracts on
  Human Factors in Computing Systems}} (Portland, OR, USA)
  \emph{(\bibinfo{series}{CHI EA '05})}. \bibinfo{publisher}{Association for
  Computing Machinery}, \bibinfo{address}{New York, NY, USA},
  \bibinfo{pages}{1945–1948}.
\newblock
\showISBNx{1595930027}
\urldef\tempurl%
\url{https://doi.org/10.1145/1056808.1057063}
\showDOI{\tempurl}


\bibitem[{Jakob Nielsen}(1994)]%
        {jakob_nielsen_10_1994}
\bibfield{author}{\bibinfo{person}{{Jakob Nielsen}}.}
  \bibinfo{year}{1994}\natexlab{}.
\newblock \bibinfo{title}{10 {Usability} {Heuristics} for {User} {Interface}
  {Design}}.
\newblock
\newblock
\urldef\tempurl%
\url{https://www.nngroup.com/articles/ten-usability-heuristics/}
\showURL{%
\tempurl}


\bibitem[Javed et~al\mbox{.}(2019)]%
        {alexareliability}
\bibfield{author}{\bibinfo{person}{Yousra Javed}, \bibinfo{person}{Shashank
  Sethi}, {and} \bibinfo{person}{Akshay Jadoun}.}
  \bibinfo{year}{2019}\natexlab{}.
\newblock \showarticletitle{Alexa's Voice Recording Behavior: A Survey of User
  Understanding and Awareness}. In \bibinfo{booktitle}{\emph{Proceedings of the
  14th International Conference on Availability, Reliability and Security}}
  (Canterbury, CA, United Kingdom) \emph{(\bibinfo{series}{ARES '19})}.
  \bibinfo{publisher}{Association for Computing Machinery},
  \bibinfo{address}{New York, NY, USA}, Article \bibinfo{articleno}{89},
  \bibinfo{numpages}{10}~pages.
\newblock
\showISBNx{9781450371643}
\urldef\tempurl%
\url{https://doi.org/10.1145/3339252.3340330}
\showDOI{\tempurl}


\bibitem[Jin et~al\mbox{.}(2022)]%
        {jin2022exploring}
\bibfield{author}{\bibinfo{person}{Haojian Jin}, \bibinfo{person}{Boyuan Guo},
  \bibinfo{person}{Rituparna Roychoudhury}, \bibinfo{person}{Yaxing Yao},
  \bibinfo{person}{Swarun Kumar}, \bibinfo{person}{Yuvraj Agarwal}, {and}
  \bibinfo{person}{Jason~I Hong}.} \bibinfo{year}{2022}\natexlab{}.
\newblock \showarticletitle{Exploring the needs of users for supporting
  privacy-protective behaviors in smart homes}. In
  \bibinfo{booktitle}{\emph{Proceedings of the 2022 CHI Conference on Human
  Factors in Computing Systems}}. \bibinfo{pages}{1--19}.
\newblock


\bibitem[Khemani and Reeves(2022)]%
        {10.1145/3491102.3517623}
\bibfield{author}{\bibinfo{person}{Krishika~Haresh Khemani} {and}
  \bibinfo{person}{Stuart Reeves}.} \bibinfo{year}{2022}\natexlab{}.
\newblock \showarticletitle{Unpacking Practitioners’ Attitudes Towards
  Codifications of Design Knowledge for Voice User Interfaces}. In
  \bibinfo{booktitle}{\emph{Proceedings of the 2022 CHI Conference on Human
  Factors in Computing Systems}} (New Orleans, LA, USA)
  \emph{(\bibinfo{series}{CHI '22})}. \bibinfo{publisher}{Association for
  Computing Machinery}, \bibinfo{address}{New York, NY, USA}, Article
  \bibinfo{articleno}{55}, \bibinfo{numpages}{10}~pages.
\newblock
\showISBNx{9781450391573}
\urldef\tempurl%
\url{https://doi.org/10.1145/3491102.3517623}
\showDOI{\tempurl}


\bibitem[Kim(2021)]%
        {kim_exploring_2021}
\bibfield{author}{\bibinfo{person}{Sunyoung Kim}.}
  \bibinfo{year}{2021}\natexlab{}.
\newblock \showarticletitle{Exploring {How} {Older} {Adults} {Use} a {Smart}
  {Speaker}–{Based} {Voice} {Assistant} in {Their} {First} {Interactions}:
  {Qualitative} {Study}}.
\newblock \bibinfo{journal}{\emph{JMIR mHealth and uHealth}}
  \bibinfo{volume}{9}, \bibinfo{number}{1} (\bibinfo{date}{Jan.}
  \bibinfo{year}{2021}), \bibinfo{pages}{e20427}.
\newblock
\showISSN{2291-5222}
\urldef\tempurl%
\url{https://doi.org/10.2196/20427}
\showDOI{\tempurl}


\bibitem[Kim et~al\mbox{.}(2021)]%
        {kim_designers_2021}
\bibfield{author}{\bibinfo{person}{Yelim Kim}, \bibinfo{person}{Mohi Reza},
  \bibinfo{person}{Joanna McGrenere}, {and} \bibinfo{person}{Dongwook Yoon}.}
  \bibinfo{year}{2021}\natexlab{}.
\newblock \showarticletitle{Designers {Characterize} {Naturalness} in {Voice}
  {User} {Interfaces}: {Their} {Goals}, {Practices}, and {Challenges}}. In
  \bibinfo{booktitle}{\emph{Proceedings of the 2021 {CHI} {Conference} on
  {Human} {Factors} in {Computing} {Systems}}} \emph{(\bibinfo{series}{{CHI}
  '21})}. \bibinfo{publisher}{Association for Computing Machinery},
  \bibinfo{address}{New York, NY, USA}, \bibinfo{pages}{1--13}.
\newblock
\showISBNx{978-1-4503-8096-6}
\urldef\tempurl%
\url{https://doi.org/10.1145/3411764.3445579}
\showDOI{\tempurl}


\bibitem[Kingaby(2022)]%
        {kingaby_voice_2022}
\bibfield{author}{\bibinfo{person}{Simon~A. Kingaby}.}
  \bibinfo{year}{2022}\natexlab{}.
\newblock \showarticletitle{Voice {User} {Interfaces}}.
\newblock In \bibinfo{booktitle}{\emph{Data-{Driven} {Alexa} {Skills}: {Voice}
  {Access} to {Rich} {Data} {Sources} for {Enterprise} {Applications}}},
  \bibfield{editor}{\bibinfo{person}{Simon~A. Kingaby}} (Ed.).
  \bibinfo{publisher}{Apress}, \bibinfo{address}{Berkeley, CA},
  \bibinfo{pages}{3--14}.
\newblock
\showISBNx{978-1-4842-7449-1}
\urldef\tempurl%
\url{https://doi.org/10.1007/978-1-4842-7449-1_1}
\showDOI{\tempurl}


\bibitem[Kučera(2017)]%
        {kucera_probing_2017}
\bibfield{author}{\bibinfo{person}{James Scott~Jan Kučera}.}
  \bibinfo{year}{2017}\natexlab{}.
\newblock \showarticletitle{Probing calmness in applications using a calm
  display prototype†}.
\newblock  (\bibinfo{year}{2017}), \bibinfo{pages}{5}.
\newblock


\bibitem[Lau et~al\mbox{.}(2016)]%
        {lau_alexa_nodate}
\bibfield{author}{\bibinfo{person}{Josephine Lau}, \bibinfo{person}{Benjamin
  Zimmerman}, {and} \bibinfo{person}{Florian Schaub}.}
  \bibinfo{year}{2016}\natexlab{}.
\newblock \showarticletitle{``{Alexa}, {Stop} {Recording}'': {Mismatches}
  between {Smart} {Speaker} {Privacy} {Controls} and {User} {Needs}}.
\newblock  (\bibinfo{year}{2016}), \bibinfo{pages}{6}.
\newblock


\bibitem[Lau et~al\mbox{.}(2018)]%
        {lau2018alexa}
\bibfield{author}{\bibinfo{person}{Josephine Lau}, \bibinfo{person}{Benjamin
  Zimmerman}, {and} \bibinfo{person}{Florian Schaub}.}
  \bibinfo{year}{2018}\natexlab{}.
\newblock \showarticletitle{Alexa, are you listening? Privacy perceptions,
  concerns and privacy-seeking behaviors with smart speakers}.
\newblock \bibinfo{journal}{\emph{Proceedings of the ACM on human-computer
  interaction}} \bibinfo{volume}{2}, \bibinfo{number}{CSCW}
  (\bibinfo{year}{2018}), \bibinfo{pages}{1--31}.
\newblock


\bibitem[Lee et~al\mbox{.}(2016)]%
        {lee_son_kim_2016}
\bibfield{author}{\bibinfo{person}{Ae~Ri Lee}, \bibinfo{person}{Soo-Min Son},
  {and} \bibinfo{person}{Kyung~Kyu Kim}.} \bibinfo{year}{2016}\natexlab{}.
\newblock \showarticletitle{Information and communication technology overload
  and social networking service fatigue: A stress perspective}.
\newblock \bibinfo{journal}{\emph{Computers in Human Behavior}}
  \bibinfo{volume}{55} (\bibinfo{year}{2016}), \bibinfo{pages}{51–61}.
\newblock
\urldef\tempurl%
\url{https://doi.org/10.1016/j.chb.2015.08.011}
\showDOI{\tempurl}


\bibitem[Lefebvre(2009)]%
        {accessphone}
\bibfield{author}{\bibinfo{person}{Craig Lefebvre}.}
  \bibinfo{year}{2009}\natexlab{}.
\newblock \showarticletitle{Integrating Cell Phones and Mobile Technologies
  Into Public Health Practice: A Social Marketing Perspective}.
\newblock \bibinfo{journal}{\emph{Health Promotion Practice}}
  \bibinfo{volume}{10}, \bibinfo{number}{4} (\bibinfo{year}{2009}),
  \bibinfo{pages}{490--494}.
\newblock
\urldef\tempurl%
\url{https://doi.org/10.1177/1524839909342849}
\showDOI{\tempurl}
\showeprint{https://doi.org/10.1177/1524839909342849}
\newblock
\shownote{PMID: 19809002}.


\bibitem[Liew et~al\mbox{.}(2022)]%
        {liew2022alexa}
\bibfield{author}{\bibinfo{person}{Tze~Wei Liew}, \bibinfo{person}{Su-Mae Tan},
  \bibinfo{person}{Wei~Ming Pang}, \bibinfo{person}{Mohammad Tariqul~Islam
  Khan}, {and} \bibinfo{person}{Si~Na Kew}.} \bibinfo{year}{2022}\natexlab{}.
\newblock \showarticletitle{I am Alexa, your virtual tutor!: The effects of
  Amazon Alexa’s text-to-speech voice enthusiasm in a multimedia learning
  environment}.
\newblock \bibinfo{journal}{\emph{Education and Information Technologies}}
  (\bibinfo{year}{2022}), \bibinfo{pages}{1--35}.
\newblock


\bibitem[Lim et~al\mbox{.}(2022)]%
        {lim2022alexa}
\bibfield{author}{\bibinfo{person}{Weng~Marc Lim}, \bibinfo{person}{Satish
  Kumar}, \bibinfo{person}{Sanjeev Verma}, {and} \bibinfo{person}{Rijul
  Chaturvedi}.} \bibinfo{year}{2022}\natexlab{}.
\newblock \showarticletitle{Alexa, what do we know about conversational
  commerce? Insights from a systematic literature review}.
\newblock \bibinfo{journal}{\emph{Psychology \& Marketing}}
  \bibinfo{volume}{39}, \bibinfo{number}{6} (\bibinfo{year}{2022}),
  \bibinfo{pages}{1129--1155}.
\newblock


\bibitem[Lopatovska et~al\mbox{.}(2019)]%
        {lopatovska_talk_2019}
\bibfield{author}{\bibinfo{person}{Irene Lopatovska}, \bibinfo{person}{Katrina
  Rink}, \bibinfo{person}{Ian Knight}, \bibinfo{person}{Kieran Raines},
  \bibinfo{person}{Kevin Cosenza}, \bibinfo{person}{Harriet Williams},
  \bibinfo{person}{Perachya Sorsche}, \bibinfo{person}{David Hirsch},
  \bibinfo{person}{Qi Li}, {and} \bibinfo{person}{Adrianna Martinez}.}
  \bibinfo{year}{2019}\natexlab{}.
\newblock \showarticletitle{Talk to me: {Exploring} user interactions with the
  {Amazon} {Alexa}}.
\newblock \bibinfo{journal}{\emph{Journal of Librarianship and Information
  Science}} \bibinfo{volume}{51}, \bibinfo{number}{4} (\bibinfo{date}{Dec.}
  \bibinfo{year}{2019}), \bibinfo{pages}{984--997}.
\newblock
\showISSN{0961-0006}
\urldef\tempurl%
\url{https://doi.org/10.1177/0961000618759414}
\showDOI{\tempurl}
\newblock
\shownote{Publisher: SAGE Publications Ltd}.


\bibitem[Luger and Sellen(2016)]%
        {luger_like_2016}
\bibfield{author}{\bibinfo{person}{Ewa Luger} {and} \bibinfo{person}{Abigail
  Sellen}.} \bibinfo{year}{2016}\natexlab{}.
\newblock \showarticletitle{"{Like} {Having} a {Really} {Bad} {PA}": {The}
  {Gulf} between {User} {Expectation} and {Experience} of {Conversational}
  {Agents}}. In \bibinfo{booktitle}{\emph{Proceedings of the 2016 {CHI}
  {Conference} on {Human} {Factors} in {Computing} {Systems}}}
  \emph{(\bibinfo{series}{{CHI} '16})}. \bibinfo{publisher}{Association for
  Computing Machinery}, \bibinfo{address}{New York, NY, USA},
  \bibinfo{pages}{5286--5297}.
\newblock
\showISBNx{978-1-4503-3362-7}
\urldef\tempurl%
\url{https://doi.org/10.1145/2858036.2858288}
\showDOI{\tempurl}


\bibitem[Lukava et~al\mbox{.}(2022)]%
        {lukava_morgado_ramirez_barbareschi_2022}
\bibfield{author}{\bibinfo{person}{Tamari Lukava},
  \bibinfo{person}{Dafne~Zuleima Morgado~Ramirez}, {and}
  \bibinfo{person}{Giulia Barbareschi}.} \bibinfo{year}{2022}\natexlab{}.
\newblock \showarticletitle{Two sides of the same coin: Accessibility practices
  and neurodivergent users' experience of extended reality}.
\newblock \bibinfo{journal}{\emph{Journal of Enabling Technologies}}
  \bibinfo{volume}{16}, \bibinfo{number}{2} (\bibinfo{year}{2022}),
  \bibinfo{pages}{75–90}.
\newblock
\urldef\tempurl%
\url{https://doi.org/10.1108/jet-03-2022-0025}
\showDOI{\tempurl}


\bibitem[Luo et~al\mbox{.}(2020)]%
        {luo_tandemtrack_2020}
\bibfield{author}{\bibinfo{person}{Yuhan Luo}, \bibinfo{person}{Bongshin Lee},
  {and} \bibinfo{person}{Eun~Kyoung Choe}.} \bibinfo{year}{2020}\natexlab{}.
\newblock \showarticletitle{{TandemTrack}: {Shaping} {Consistent} {Exercise}
  {Experience} by {Complementing} a {Mobile} {App} with a {Smart} {Speaker}}.
  In \bibinfo{booktitle}{\emph{Proceedings of the 2020 {CHI} {Conference} on
  {Human} {Factors} in {Computing} {Systems}}} \emph{(\bibinfo{series}{{CHI}
  '20})}. \bibinfo{publisher}{Association for Computing Machinery},
  \bibinfo{address}{New York, NY, USA}, \bibinfo{pages}{1--13}.
\newblock
\showISBNx{978-1-4503-6708-0}
\urldef\tempurl%
\url{https://doi.org/10.1145/3313831.3376616}
\showDOI{\tempurl}


\bibitem[Luria et~al\mbox{.}(2017)]%
        {luria}
\bibfield{author}{\bibinfo{person}{Michal Luria}, \bibinfo{person}{Guy
  Hoffman}, {and} \bibinfo{person}{Oren Zuckerman}.}
  \bibinfo{year}{2017}\natexlab{}.
\newblock \showarticletitle{Comparing Social Robot, Screen and Voice Interfaces
  for Smart-Home Control}. In \bibinfo{booktitle}{\emph{Proceedings of the 2017
  CHI Conference on Human Factors in Computing Systems}} (Denver, Colorado,
  USA) \emph{(\bibinfo{series}{CHI '17})}. \bibinfo{publisher}{Association for
  Computing Machinery}, \bibinfo{address}{New York, NY, USA},
  \bibinfo{pages}{580–628}.
\newblock
\showISBNx{9781450346559}
\urldef\tempurl%
\url{https://doi.org/10.1145/3025453.3025786}
\showDOI{\tempurl}


\bibitem[López et~al\mbox{.}(2018)]%
        {lopez_alexa_2018}
\bibfield{author}{\bibinfo{person}{Gustavo López}, \bibinfo{person}{Luis
  Quesada}, {and} \bibinfo{person}{Luis~A. Guerrero}.}
  \bibinfo{year}{2018}\natexlab{}.
\newblock \showarticletitle{Alexa vs. {Siri} vs. {Cortana} vs. {Google}
  {Assistant}: {A} {Comparison} of {Speech}-{Based} {Natural} {User}
  {Interfaces}}. In \bibinfo{booktitle}{\emph{Advances in {Human} {Factors} and
  {Systems} {Interaction}}} \emph{(\bibinfo{series}{Advances in {Intelligent}
  {Systems} and {Computing}})}, \bibfield{editor}{\bibinfo{person}{Isabel~L.
  Nunes}} (Ed.). \bibinfo{publisher}{Springer International Publishing},
  \bibinfo{address}{Cham}, \bibinfo{pages}{241--250}.
\newblock
\showISBNx{978-3-319-60366-7}
\urldef\tempurl%
\url{https://doi.org/10.1007/978-3-319-60366-7_23}
\showDOI{\tempurl}


\bibitem[Maguire(2019)]%
        {maguire_development_2019}
\bibfield{author}{\bibinfo{person}{Martin Maguire}.}
  \bibinfo{year}{2019}\natexlab{}.
\newblock \showarticletitle{Development of a {Heuristic} {Evaluation} {Tool}
  for {Voice} {User} {Interfaces}}. In \bibinfo{booktitle}{\emph{Design, {User}
  {Experience}, and {Usability}. {Practice} and {Case} {Studies}}}
  \emph{(\bibinfo{series}{Lecture {Notes} in {Computer} {Science}})},
  \bibfield{editor}{\bibinfo{person}{Aaron Marcus} {and}
  \bibinfo{person}{Wentao Wang}} (Eds.). \bibinfo{publisher}{Springer
  International Publishing}, \bibinfo{address}{Cham},
  \bibinfo{pages}{212--225}.
\newblock
\showISBNx{978-3-030-23535-2}
\urldef\tempurl%
\url{https://doi.org/10.1007/978-3-030-23535-2_16}
\showDOI{\tempurl}


\bibitem[Major et~al\mbox{.}(2021)]%
        {majordavid_alexa_2021}
\bibfield{author}{\bibinfo{person}{David Major}, \bibinfo{person}{Danny~Yuxing
  Huang}, \bibinfo{person}{Marshini Chetty}, {and} \bibinfo{person}{Nick
  Feamster}.} \bibinfo{year}{2021}\natexlab{}.
\newblock \showarticletitle{Alexa, {Who} {Am} {I} {Speaking} {To}?:
  {Understanding} {Users}’ {Ability} to {Identify} {Third}-{Party} {Apps} on
  {Amazon} {Alexa}}.
\newblock \bibinfo{journal}{\emph{ACM Transactions on Internet Technology
  (TOIT)}} (\bibinfo{date}{Sept.} \bibinfo{year}{2021}).
\newblock
\urldef\tempurl%
\url{https://doi.org/10.1145/3446389}
\showDOI{\tempurl}
\newblock
\shownote{Publisher: ACM PUB27 New York, NY}.


\bibitem[Malkin et~al\mbox{.}(2019)]%
        {alwayslisten1}
\bibfield{author}{\bibinfo{person}{Nathan Malkin}, \bibinfo{person}{Serge
  Egelman}, {and} \bibinfo{person}{David Wagner}.}
  \bibinfo{year}{2019}\natexlab{}.
\newblock \showarticletitle{Privacy Controls for Always-Listening Devices}. In
  \bibinfo{booktitle}{\emph{Proceedings of the New Security Paradigms
  Workshop}} (San Carlos, Costa Rica) \emph{(\bibinfo{series}{NSPW '19})}.
  \bibinfo{publisher}{Association for Computing Machinery},
  \bibinfo{address}{New York, NY, USA}, \bibinfo{pages}{78–91}.
\newblock
\showISBNx{9781450376471}
\urldef\tempurl%
\url{https://doi.org/10.1145/3368860.3368867}
\showDOI{\tempurl}


\bibitem[Malodia et~al\mbox{.}(2022)]%
        {malodia2022can}
\bibfield{author}{\bibinfo{person}{Suresh Malodia}, \bibinfo{person}{Alberto
  Ferraris}, \bibinfo{person}{Mototaka Sakashita}, \bibinfo{person}{Amandeep
  Dhir}, {and} \bibinfo{person}{Beata Gavurova}.}
  \bibinfo{year}{2022}\natexlab{}.
\newblock \showarticletitle{Can Alexa serve customers better? AI-driven voice
  assistant service interactions}.
\newblock \bibinfo{journal}{\emph{Journal of Services Marketing}}
  \bibinfo{number}{ahead-of-print} (\bibinfo{year}{2022}).
\newblock


\bibitem[Mande et~al\mbox{.}(2021)]%
        {mande_deaf_2021}
\bibfield{author}{\bibinfo{person}{Vaishnavi Mande}, \bibinfo{person}{Abraham
  Glasser}, \bibinfo{person}{Becca Dingman}, {and} \bibinfo{person}{Matt
  Huenerfauth}.} \bibinfo{year}{2021}\natexlab{}.
\newblock \showarticletitle{Deaf {Users}’ {Preferences} {Among} {Wake}-{Up}
  {Approaches} during {Sign}-{Language} {Interaction} with {Personal}
  {Assistant} {Devices}}. In \bibinfo{booktitle}{\emph{Extended {Abstracts} of
  the 2021 {CHI} {Conference} on {Human} {Factors} in {Computing} {Systems}}}
  \emph{(\bibinfo{series}{{CHI} {EA} '21})}. \bibinfo{publisher}{Association
  for Computing Machinery}, \bibinfo{address}{New York, NY, USA},
  \bibinfo{pages}{1--6}.
\newblock
\showISBNx{978-1-4503-8095-9}
\urldef\tempurl%
\url{https://doi.org/10.1145/3411763.3451592}
\showDOI{\tempurl}


\bibitem[Manresa-Yee et~al\mbox{.}(2010)]%
        {manresa_yee_ponsa_varona_perales_2010}
\bibfield{author}{\bibinfo{person}{Cristina Manresa-Yee}, \bibinfo{person}{Pere
  Ponsa}, \bibinfo{person}{Javier Varona}, {and} \bibinfo{person}{Francisco~J.
  Perales}.} \bibinfo{year}{2010}\natexlab{}.
\newblock \showarticletitle{User experience to improve the usability of a
  vision-based interface}.
\newblock \bibinfo{journal}{\emph{Interacting with Computers}}
  \bibinfo{volume}{22}, \bibinfo{number}{6} (\bibinfo{year}{2010}),
  \bibinfo{pages}{594–605}.
\newblock
\urldef\tempurl%
\url{https://doi.org/10.1016/j.intcom.2010.06.004}
\showDOI{\tempurl}


\bibitem[Marikyan et~al\mbox{.}(2022)]%
        {marikyan2022alexa}
\bibfield{author}{\bibinfo{person}{Davit Marikyan}, \bibinfo{person}{Savvas
  Papagiannidis}, \bibinfo{person}{Omer~F Rana}, \bibinfo{person}{Rajiv
  Ranjan}, {and} \bibinfo{person}{Graham Morgan}.}
  \bibinfo{year}{2022}\natexlab{}.
\newblock \showarticletitle{“Alexa, let’s talk about my productivity”:
  The impact of digital assistants on work productivity}.
\newblock \bibinfo{journal}{\emph{Journal of Business Research}}
  \bibinfo{volume}{142} (\bibinfo{year}{2022}), \bibinfo{pages}{572--584}.
\newblock


\bibitem[Matthes et~al\mbox{.}(2020)]%
        {matthes_too_2020}
\bibfield{author}{\bibinfo{person}{Jörg Matthes}, \bibinfo{person}{Kathrin
  Karsay}, \bibinfo{person}{Desirée Schmuck}, {and} \bibinfo{person}{Anja
  Stevic}.} \bibinfo{year}{2020}\natexlab{}.
\newblock \showarticletitle{“{Too} much to handle”: {Impact} of mobile
  social networking sites on information overload, depressive symptoms, and
  well-being}.
\newblock \bibinfo{journal}{\emph{Computers in Human Behavior}}
  \bibinfo{volume}{105} (\bibinfo{date}{April} \bibinfo{year}{2020}),
  \bibinfo{pages}{106217}.
\newblock
\showISSN{0747-5632}
\urldef\tempurl%
\url{https://doi.org/10.1016/j.chb.2019.106217}
\showDOI{\tempurl}


\bibitem[Mavrina et~al\mbox{.}(2022)]%
        {mavrina_alexa_2022}
\bibfield{author}{\bibinfo{person}{Lina Mavrina}, \bibinfo{person}{Jessica
  Szczuka}, \bibinfo{person}{Clara Strathmann}, \bibinfo{person}{Lisa~Michelle
  Bohnenkamp}, \bibinfo{person}{Nicole Krämer}, {and} \bibinfo{person}{Stefan
  Kopp}.} \bibinfo{year}{2022}\natexlab{}.
\newblock \showarticletitle{“{Alexa}, {You}'re {Really} {Stupid}”: {A}
  {Longitudinal} {Field} {Study} on {Communication} {Breakdowns} {Between}
  {Family} {Members} and a {Voice} {Assistant}}.
\newblock \bibinfo{journal}{\emph{Frontiers in Computer Science}}
  \bibinfo{volume}{4} (\bibinfo{date}{Jan.} \bibinfo{year}{2022}),
  \bibinfo{pages}{791704}.
\newblock
\showISSN{2624-9898}
\urldef\tempurl%
\url{https://doi.org/10.3389/fcomp.2022.791704}
\showDOI{\tempurl}


\bibitem[Murad et~al\mbox{.}(2018)]%
        {gapincalmminimalist}
\bibfield{author}{\bibinfo{person}{Christine Murad}, \bibinfo{person}{Cosmin
  Munteanu}, \bibinfo{person}{Leigh Clark}, {and} \bibinfo{person}{Benjamin~R.
  Cowan}.} \bibinfo{year}{2018}\natexlab{}.
\newblock \showarticletitle{Design Guidelines for Hands-Free Speech
  Interaction}. In \bibinfo{booktitle}{\emph{Proceedings of the 20th
  International Conference on Human-Computer Interaction with Mobile Devices
  and Services Adjunct}} (Barcelona, Spain) \emph{(\bibinfo{series}{MobileHCI
  '18})}. \bibinfo{publisher}{Association for Computing Machinery},
  \bibinfo{address}{New York, NY, USA}, \bibinfo{pages}{269–276}.
\newblock
\showISBNx{9781450359412}
\urldef\tempurl%
\url{https://doi.org/10.1145/3236112.3236149}
\showDOI{\tempurl}


\bibitem[Murad et~al\mbox{.}(2019)]%
        {murad_revolution_2019}
\bibfield{author}{\bibinfo{person}{Christine Murad}, \bibinfo{person}{Cosmin
  Munteanu}, \bibinfo{person}{Benjamin~R. Cowan}, {and} \bibinfo{person}{Leigh
  Clark}.} \bibinfo{year}{2019}\natexlab{}.
\newblock \showarticletitle{Revolution or {Evolution}? {Speech} {Interaction}
  and {HCI} {Design} {Guidelines}}.
\newblock \bibinfo{journal}{\emph{IEEE Pervasive Computing}}
  \bibinfo{volume}{18}, \bibinfo{number}{2} (\bibinfo{date}{April}
  \bibinfo{year}{2019}), \bibinfo{pages}{33--45}.
\newblock
\showISSN{1558-2590}
\urldef\tempurl%
\url{https://doi.org/10.1109/MPRV.2019.2906991}
\showDOI{\tempurl}
\newblock
\shownote{Conference Name: IEEE Pervasive Computing}.


\bibitem[Murad et~al\mbox{.}(2021)]%
        {murad_finding_2021}
\bibfield{author}{\bibinfo{person}{Christine Murad}, \bibinfo{person}{Cosmin
  Munteanu}, \bibinfo{person}{Benjamin R.~Cowan}, {and} \bibinfo{person}{Leigh
  Clark}.} \bibinfo{year}{2021}\natexlab{}.
\newblock \showarticletitle{Finding a {New} {Voice}: {Transitioning}
  {Designers} from {GUI} to {VUI} {Design}}. In \bibinfo{booktitle}{\emph{{CUI}
  2021 - 3rd {Conference} on {Conversational} {User} {Interfaces}}}.
  \bibinfo{publisher}{ACM}, \bibinfo{address}{Bilbao (online) Spain},
  \bibinfo{pages}{1--12}.
\newblock
\showISBNx{978-1-4503-8998-3}
\urldef\tempurl%
\url{https://doi.org/10.1145/3469595.3469617}
\showDOI{\tempurl}


\bibitem[Mutchler(2018)]%
        {mutchler_2018}
\bibfield{author}{\bibinfo{person}{Ava Mutchler}.}
  \bibinfo{year}{2018}\natexlab{}.
\newblock \bibinfo{title}{A timeline of voice assistant and Smart Speaker
  Technology from 1961 to today}.
\newblock
\newblock
\urldef\tempurl%
\url{https://voicebot.ai/2018/03/28/timeline-voice-assistant-smart-speaker-technology-1961-today/}
\showURL{%
\tempurl}


\bibitem[{National Public Radio (NPR)}(2020)]%
        {national_public_radio_npr_use_2020}
\bibfield{author}{\bibinfo{person}{{National Public Radio (NPR)}}.}
  \bibinfo{year}{2020}\natexlab{}.
\newblock \showarticletitle{Use of {Smart} {Speakers} in the {U}.{S}.
  {Increases} {During} {Quarantine}}.
\newblock \bibinfo{journal}{\emph{NPR}} (\bibinfo{date}{April}
  \bibinfo{year}{2020}).
\newblock
\urldef\tempurl%
\url{https://www.npr.org/about-npr/848319916/use-of-smart-speakers-in-the-u-s-increases-during-quarantine}
\showURL{%
\tempurl}


\bibitem[{National Public Radio (NPR)}(2022)]%
        {national_public_radio_npr_npr_2022}
\bibfield{author}{\bibinfo{person}{{National Public Radio (NPR)}}.}
  \bibinfo{year}{2022}\natexlab{}.
\newblock \showarticletitle{{NPR} \& {Edison} {Research}: {Smart} {Speaker}
  {Ownership} {Reaches} 35\% of {Americans}}.
\newblock \bibinfo{journal}{\emph{NPR}} (\bibinfo{date}{June}
  \bibinfo{year}{2022}).
\newblock
\urldef\tempurl%
\url{https://www.npr.org/about-npr/1105579648/npr-edison-research-smart-speaker-ownership-reaches-35-of-americans}
\showURL{%
\tempurl}


\bibitem[Ning et~al\mbox{.}(2019)]%
        {companies_release}
\bibfield{author}{\bibinfo{person}{Yishuang Ning}, \bibinfo{person}{Sheng He},
  \bibinfo{person}{Chunxiao Xing}, {and} \bibinfo{person}{Liang-Jie Zhang}.}
  \bibinfo{year}{2019}\natexlab{}.
\newblock \showarticletitle{The Development Trend of Intelligent Speech
  Interaction}. In \bibinfo{booktitle}{\emph{Cognitive Computing -- ICCC
  2019}}, \bibfield{editor}{\bibinfo{person}{Ruifeng Xu},
  \bibinfo{person}{Jianzong Wang}, {and} \bibinfo{person}{Liang-Jie Zhang}}
  (Eds.). \bibinfo{publisher}{Springer International Publishing},
  \bibinfo{address}{Cham}, \bibinfo{pages}{169--179}.
\newblock
\showISBNx{978-3-030-23407-2}


\bibitem[Norman(1983)]%
        {norman_observations_1983}
\bibfield{author}{\bibinfo{person}{Don Norman}.}
  \bibinfo{year}{1983}\natexlab{}.
\newblock \bibinfo{booktitle}{\emph{Some {Observations} on {Mental} {Models}}}.
\newblock \bibinfo{publisher}{Psychology Press}.
\newblock
\showISBNx{978-1-315-80272-5}
\urldef\tempurl%
\url{https://doi.org/10.4324/9781315802725-5}
\showDOI{\tempurl}
\newblock
\shownote{Pages: 15-22 Publication Title: Mental Models}.


\bibitem[Nowacki et~al\mbox{.}(2020)]%
        {nowacki_improving_2020}
\bibfield{author}{\bibinfo{person}{Caroline Nowacki}, \bibinfo{person}{Anna
  Gordeeva}, {and} \bibinfo{person}{Anne-Hélène Lizé}.}
  \bibinfo{year}{2020}\natexlab{}.
\newblock \showarticletitle{Improving the {Usability} of {Voice} {User}
  {Interfaces}: {A} {New} {Set} of {Ergonomic} {Criteria}}. In
  \bibinfo{booktitle}{\emph{Design, {User} {Experience}, and {Usability}.
  {Design} for {Contemporary} {Interactive} {Environments}}}
  \emph{(\bibinfo{series}{Lecture {Notes} in {Computer} {Science}})},
  \bibfield{editor}{\bibinfo{person}{Aaron Marcus} {and}
  \bibinfo{person}{Elizabeth Rosenzweig}} (Eds.). \bibinfo{publisher}{Springer
  International Publishing}, \bibinfo{address}{Cham},
  \bibinfo{pages}{117--133}.
\newblock
\showISBNx{978-3-030-49760-6}
\urldef\tempurl%
\url{https://doi.org/10.1007/978-3-030-49760-6_8}
\showDOI{\tempurl}


\bibitem[Occhialini et~al\mbox{.}(2011)]%
        {ambient_2011}
\bibfield{author}{\bibinfo{person}{Valentina Occhialini}, \bibinfo{person}{Harm
  van Essen}, {and} \bibinfo{person}{Berry Eggen}.}
  \bibinfo{year}{2011}\natexlab{}.
\newblock \showarticletitle{Design and evaluation of an ambient display to
  support time management during meetings}.
\newblock \bibinfo{journal}{\emph{Human-Computer Interaction – INTERACT
  2011}} (\bibinfo{date}{Jan} \bibinfo{year}{2011}),
  \bibinfo{pages}{263–280}.
\newblock
\urldef\tempurl%
\url{https://doi.org/10.1007/978-3-642-23771-3_20}
\showDOI{\tempurl}


\bibitem[Olmstead(2017)]%
        {olmstead_nearly_nodate}
\bibfield{author}{\bibinfo{person}{Kenneth Olmstead}.}
  \bibinfo{year}{2017}\natexlab{}.
\newblock \bibinfo{title}{Nearly half of {Americans} use digital voice
  assistants, mostly on their smartphones}.
\newblock
\newblock
\urldef\tempurl%
\url{https://www.pewresearch.org/fact-tank/2017/12/12/nearly-half-of-americans-use-digital-voice-assistants-mostly-on-their-smartphones/}
\showURL{%
\tempurl}


\bibitem[Orlofsky and Wozniak(2022)]%
        {orlofsky2022older}
\bibfield{author}{\bibinfo{person}{Sarit Orlofsky} {and}
  \bibinfo{person}{Kathryn Wozniak}.} \bibinfo{year}{2022}\natexlab{}.
\newblock \showarticletitle{Older adults' experiences using Alexa}.
\newblock \bibinfo{journal}{\emph{Geriatric Nursing}}  \bibinfo{volume}{48}
  (\bibinfo{year}{2022}), \bibinfo{pages}{240--250}.
\newblock


\bibitem[Ostrowski et~al\mbox{.}(2022)]%
        {mentalmodelemotion}
\bibfield{author}{\bibinfo{person}{Anastasia~K. Ostrowski},
  \bibinfo{person}{Jenny Fu}, \bibinfo{person}{Vasiliki Zygouras},
  \bibinfo{person}{Hae~Won Park}, {and} \bibinfo{person}{Cynthia Breazeal}.}
  \bibinfo{year}{2022}\natexlab{}.
\newblock \showarticletitle{Speed Dating with Voice User Interfaces:
  Understanding How Families Interact and Perceive Voice User Interfaces in a
  Group Setting}.
\newblock \bibinfo{journal}{\emph{Frontiers in Robotics and AI}}
  \bibinfo{volume}{8} (\bibinfo{year}{2022}).
\newblock
\showISSN{2296-9144}
\urldef\tempurl%
\url{https://doi.org/10.3389/frobt.2021.730992}
\showDOI{\tempurl}


\bibitem[Pan et~al\mbox{.}(2022)]%
        {commandcus}
\bibfield{author}{\bibinfo{person}{Lihang Pan}, \bibinfo{person}{Chun Yu},
  \bibinfo{person}{JiaHui Li}, \bibinfo{person}{Tian Huang},
  \bibinfo{person}{Xiaojun Bi}, {and} \bibinfo{person}{Yuanchun Shi}.}
  \bibinfo{year}{2022}\natexlab{}.
\newblock \showarticletitle{Automatically Generating and Improving Voice
  Command Interface from Operation Sequences on Smartphones}. In
  \bibinfo{booktitle}{\emph{Proceedings of the 2022 CHI Conference on Human
  Factors in Computing Systems}} (New Orleans, LA, USA)
  \emph{(\bibinfo{series}{CHI '22})}. \bibinfo{publisher}{Association for
  Computing Machinery}, \bibinfo{address}{New York, NY, USA}, Article
  \bibinfo{articleno}{208}, \bibinfo{numpages}{21}~pages.
\newblock
\showISBNx{9781450391573}
\urldef\tempurl%
\url{https://doi.org/10.1145/3491102.3517459}
\showDOI{\tempurl}


\bibitem[Parvin et~al\mbox{.}(2022)]%
        {10.1145/3531073.3531157}
\bibfield{author}{\bibinfo{person}{Parvaneh Parvin}, \bibinfo{person}{Marco
  Manca}, \bibinfo{person}{Caterina Senette}, \bibinfo{person}{Maria~Claudia
  Buzzi}, \bibinfo{person}{Marina Buzzi}, {and} \bibinfo{person}{Susanna
  Pelagatti}.} \bibinfo{year}{2022}\natexlab{}.
\newblock \showarticletitle{Alexism: ALEXa Supporting Children with AutISM in
  Their Oral Care at Home}. In \bibinfo{booktitle}{\emph{Proceedings of the
  2022 International Conference on Advanced Visual Interfaces}} (Frascati,
  Rome, Italy) \emph{(\bibinfo{series}{AVI 2022})}.
  \bibinfo{publisher}{Association for Computing Machinery},
  \bibinfo{address}{New York, NY, USA}, Article \bibinfo{articleno}{18},
  \bibinfo{numpages}{5}~pages.
\newblock
\showISBNx{9781450397193}
\urldef\tempurl%
\url{https://doi.org/10.1145/3531073.3531157}
\showDOI{\tempurl}


\bibitem[Pearl(2017)]%
        {pearl_designing_nodate}
\bibfield{author}{\bibinfo{person}{Cathy Pearl}.}
  \bibinfo{year}{2017}\natexlab{}.
\newblock \showarticletitle{Designing {Voice} {User} {Interfaces}}.
\newblock  (\bibinfo{year}{2017}), \bibinfo{pages}{278}.
\newblock


\bibitem[Pfeifle(2018)]%
        {pfeifle2018alexa}
\bibfield{author}{\bibinfo{person}{Anne Pfeifle}.}
  \bibinfo{year}{2018}\natexlab{}.
\newblock \showarticletitle{Alexa, what should we do about privacy: Protecting
  privacy for users of voice-activated devices}.
\newblock \bibinfo{journal}{\emph{Wash. L. Rev.}}  \bibinfo{volume}{93}
  (\bibinfo{year}{2018}), \bibinfo{pages}{421}.
\newblock


\bibitem[Pielot and Rello(2015)]%
        {pielot_notification}
\bibfield{author}{\bibinfo{person}{Martin Pielot} {and} \bibinfo{person}{Luz
  Rello}.} \bibinfo{year}{2015}\natexlab{}.
\newblock \showarticletitle{The Do Not Disturb Challenge: A Day Without
  Notifications}. In \bibinfo{booktitle}{\emph{Proceedings of the 33rd Annual
  ACM Conference Extended Abstracts on Human Factors in Computing Systems}}
  (Seoul, Republic of Korea) \emph{(\bibinfo{series}{CHI EA '15})}.
  \bibinfo{publisher}{Association for Computing Machinery},
  \bibinfo{address}{New York, NY, USA}, \bibinfo{pages}{1761–1766}.
\newblock
\showISBNx{9781450331463}
\urldef\tempurl%
\url{https://doi.org/10.1145/2702613.2732704}
\showDOI{\tempurl}


\bibitem[Pollmann et~al\mbox{.}(2020)]%
        {pollmann2020robot}
\bibfield{author}{\bibinfo{person}{Kathrin Pollmann},
  \bibinfo{person}{Christopher Ruff}, \bibinfo{person}{Kevin Vetter}, {and}
  \bibinfo{person}{Gottfried Zimmermann}.} \bibinfo{year}{2020}\natexlab{}.
\newblock \showarticletitle{Robot vs. voice assistant: Is playing with pepper
  more fun than playing with alexa?}. In \bibinfo{booktitle}{\emph{Companion of
  the 2020 ACM/IEEE international conference on human-robot interaction}}.
  \bibinfo{pages}{395--397}.
\newblock


\bibitem[Purington et~al\mbox{.}(2017)]%
        {purington_alexa_2017}
\bibfield{author}{\bibinfo{person}{Amanda Purington},
  \bibinfo{person}{Jessie~G. Taft}, \bibinfo{person}{Shruti Sannon},
  \bibinfo{person}{Natalya~N. Bazarova}, {and} \bibinfo{person}{Samuel~Hardman
  Taylor}.} \bibinfo{year}{2017}\natexlab{}.
\newblock \showarticletitle{"{Alexa} is my new {BFF}": {Social} {Roles}, {User}
  {Satisfaction}, and {Personification} of the {Amazon} {Echo}}. In
  \bibinfo{booktitle}{\emph{Proceedings of the 2017 {CHI} {Conference}
  {Extended} {Abstracts} on {Human} {Factors} in {Computing} {Systems}}}
  \emph{(\bibinfo{series}{{CHI} {EA} '17})}. \bibinfo{publisher}{Association
  for Computing Machinery}, \bibinfo{address}{New York, NY, USA},
  \bibinfo{pages}{2853--2859}.
\newblock
\showISBNx{978-1-4503-4656-6}
\urldef\tempurl%
\url{https://doi.org/10.1145/3027063.3053246}
\showDOI{\tempurl}


\bibitem[Qiu et~al\mbox{.}(2021)]%
        {qiu_nurse_2021}
\bibfield{author}{\bibinfo{person}{Ling Qiu}, \bibinfo{person}{Bethany Kanski},
  \bibinfo{person}{Shawna Doerksen}, \bibinfo{person}{Renate Winkels},
  \bibinfo{person}{Kathryn~H Schmitz}, {and} \bibinfo{person}{Saeed Abdullah}.}
  \bibinfo{year}{2021}\natexlab{}.
\newblock \showarticletitle{Nurse {AMIE}: {Using} {Smart} {Speakers} to
  {Provide} {Supportive} {Care} {Intervention} for {Women} with {Metastatic}
  {Breast} {Cancer}}. In \bibinfo{booktitle}{\emph{Extended {Abstracts} of the
  2021 {CHI} {Conference} on {Human} {Factors} in {Computing} {Systems}}}
  \emph{(\bibinfo{series}{{CHI} {EA} '21})}. \bibinfo{publisher}{Association
  for Computing Machinery}, \bibinfo{address}{New York, NY, USA},
  \bibinfo{pages}{1--7}.
\newblock
\showISBNx{978-1-4503-8095-9}
\urldef\tempurl%
\url{https://doi.org/10.1145/3411763.3451827}
\showDOI{\tempurl}


\bibitem[Ramadan et~al\mbox{.}(2021a)]%
        {ramadan2021amazon}
\bibfield{author}{\bibinfo{person}{Zahy Ramadan}, \bibinfo{person}{Maya
  F~Farah}, {and} \bibinfo{person}{Lea El~Essrawi}.}
  \bibinfo{year}{2021}\natexlab{a}.
\newblock \showarticletitle{From Amazon. com to Amazon. love: How Alexa is
  redefining companionship and interdependence for people with special needs}.
\newblock \bibinfo{journal}{\emph{Psychology \& Marketing}}
  \bibinfo{volume}{38}, \bibinfo{number}{4} (\bibinfo{year}{2021}),
  \bibinfo{pages}{596--609}.
\newblock


\bibitem[Ramadan et~al\mbox{.}(2021b)]%
        {ramadan_amazoncom_2021}
\bibfield{author}{\bibinfo{person}{Zahy Ramadan}, \bibinfo{person}{Maya Farah},
  {and} \bibinfo{person}{Lea El~Essrawi}.} \bibinfo{year}{2021}\natexlab{b}.
\newblock \showarticletitle{From {Amazon}.com to {Amazon}.love: {How} {Alexa}
  is redefining companionship and interdependence for people with special
  needs}.
\newblock \bibinfo{journal}{\emph{Psychology \& Marketing}}
  \bibinfo{volume}{38}, \bibinfo{number}{4} (\bibinfo{date}{April}
  \bibinfo{year}{2021}), \bibinfo{pages}{596--609}.
\newblock
\showISSN{0742-6046, 1520-6793}
\urldef\tempurl%
\url{https://doi.org/10.1002/mar.21441}
\showDOI{\tempurl}


\bibitem[Reddy et~al\mbox{.}(2021)]%
        {reddy_making_2021}
\bibfield{author}{\bibinfo{person}{Anuradha Reddy}, \bibinfo{person}{A.~Baki
  Kocaballi}, \bibinfo{person}{Iohanna Nicenboim}, \bibinfo{person}{Marie
  Louise~Juul Søndergaard}, \bibinfo{person}{Maria~Luce Lupetti},
  \bibinfo{person}{Cayla Key}, \bibinfo{person}{Chris Speed},
  \bibinfo{person}{Dan Lockton}, \bibinfo{person}{Elisa Giaccardi},
  \bibinfo{person}{Francisca Grommé}, \bibinfo{person}{Holly Robbins},
  \bibinfo{person}{Namrata Primlani}, \bibinfo{person}{Paulina Yurman},
  \bibinfo{person}{Shanti Sumartojo}, \bibinfo{person}{Thao Phan},
  \bibinfo{person}{Viktor Bedö}, {and} \bibinfo{person}{Yolande Strengers}.}
  \bibinfo{year}{2021}\natexlab{}.
\newblock \showarticletitle{Making {Everyday} {Things} {Talk}: {Speculative}
  {Conversations} into the {Future} of {Voice} {Interfaces} at {Home}}. In
  \bibinfo{booktitle}{\emph{Extended {Abstracts} of the 2021 {CHI} {Conference}
  on {Human} {Factors} in {Computing} {Systems}}} \emph{(\bibinfo{series}{{CHI}
  {EA} '21})}. \bibinfo{publisher}{Association for Computing Machinery},
  \bibinfo{address}{New York, NY, USA}, \bibinfo{pages}{1--16}.
\newblock
\showISBNx{978-1-4503-8095-9}
\urldef\tempurl%
\url{https://doi.org/10.1145/3411763.3450390}
\showDOI{\tempurl}


\bibitem[Reeves et~al\mbox{.}(2018)]%
        {reeves_ml}
\bibfield{author}{\bibinfo{person}{Stuart Reeves}, \bibinfo{person}{Martin
  Porcheron}, \bibinfo{person}{Joel~E. Fischer}, \bibinfo{person}{Heloisa
  Candello}, \bibinfo{person}{Donald McMillan}, \bibinfo{person}{Moira
  McGregor}, \bibinfo{person}{Robert~J. Moore}, \bibinfo{person}{Rein
  Sikveland}, \bibinfo{person}{Alex~S. Taylor}, \bibinfo{person}{Julia
  Velkovska}, {and} \bibinfo{person}{Moustafa Zouinar}.}
  \bibinfo{year}{2018}\natexlab{}.
\newblock \showarticletitle{Voice-Based Conversational UX Studies and Design}.
  In \bibinfo{booktitle}{\emph{Extended Abstracts of the 2018 CHI Conference on
  Human Factors in Computing Systems}} (Montreal QC, Canada)
  \emph{(\bibinfo{series}{CHI EA '18})}. \bibinfo{publisher}{Association for
  Computing Machinery}, \bibinfo{address}{New York, NY, USA},
  \bibinfo{pages}{1–8}.
\newblock
\showISBNx{9781450356213}
\urldef\tempurl%
\url{https://doi.org/10.1145/3170427.3170619}
\showDOI{\tempurl}


\bibitem[{Reportlinker}(2022)]%
        {reportlinker_smart_2022}
\bibfield{author}{\bibinfo{person}{{Reportlinker}}.}
  \bibinfo{year}{2022}\natexlab{}.
\newblock \bibinfo{title}{Smart {Speakers} {Global} {Market} {Report} 2022}.
\newblock
\newblock
\urldef\tempurl%
\url{https://money.yahoo.com/smart-speakers-global-market-report-124000369.html}
\showURL{%
\tempurl}


\bibitem[Sabir et~al\mbox{.}(2022)]%
        {sabir_hey_2022}
\bibfield{author}{\bibinfo{person}{Aafaq Sabir}, \bibinfo{person}{Evan
  Lafontaine}, {and} \bibinfo{person}{Anupam Das}.}
  \bibinfo{year}{2022}\natexlab{}.
\newblock \showarticletitle{Hey {Alexa}, {Who} {Am} {I} {Talking} to?:
  {Analyzing} {Users}’ {Perception} and {Awareness} {Regarding} {Third}-party
  {Alexa} {Skills}}. In \bibinfo{booktitle}{\emph{Proceedings of the 2022 {CHI}
  {Conference} on {Human} {Factors} in {Computing} {Systems}}}
  \emph{(\bibinfo{series}{{CHI} '22})}. \bibinfo{publisher}{Association for
  Computing Machinery}, \bibinfo{address}{New York, NY, USA},
  \bibinfo{pages}{1--15}.
\newblock
\showISBNx{978-1-4503-9157-3}
\urldef\tempurl%
\url{https://doi.org/10.1145/3491102.3517510}
\showDOI{\tempurl}


\bibitem[Schoegler et~al\mbox{.}(2020)]%
        {schoegler2020use}
\bibfield{author}{\bibinfo{person}{Peter Schoegler}, \bibinfo{person}{Markus
  Ebner}, {and} \bibinfo{person}{Martin Ebner}.}
  \bibinfo{year}{2020}\natexlab{}.
\newblock \showarticletitle{The use of alexa for mass education}. In
  \bibinfo{booktitle}{\emph{EdMedia+ Innovate Learning}}. Association for the
  Advancement of Computing in Education (AACE), \bibinfo{pages}{721--730}.
\newblock


\bibitem[Sciuto et~al\mbox{.}(2018)]%
        {sciutojasonhong}
\bibfield{author}{\bibinfo{person}{Alex Sciuto}, \bibinfo{person}{Arnita
  Saini}, \bibinfo{person}{Jodi Forlizzi}, {and} \bibinfo{person}{Jason~I.
  Hong}.} \bibinfo{year}{2018}\natexlab{}.
\newblock \showarticletitle{"Hey Alexa, What's Up?": A Mixed-Methods Studies of
  In-Home Conversational Agent Usage}. In \bibinfo{booktitle}{\emph{Proceedings
  of the 2018 Designing Interactive Systems Conference}} (Hong Kong, China)
  \emph{(\bibinfo{series}{DIS '18})}. \bibinfo{publisher}{Association for
  Computing Machinery}, \bibinfo{address}{New York, NY, USA},
  \bibinfo{pages}{857–868}.
\newblock
\showISBNx{9781450351980}
\urldef\tempurl%
\url{https://doi.org/10.1145/3196709.3196772}
\showDOI{\tempurl}


\bibitem[Scott et~al\mbox{.}(2016)]%
        {scott_valley_simecka_2016}
\bibfield{author}{\bibinfo{person}{David~A. Scott}, \bibinfo{person}{Bart
  Valley}, {and} \bibinfo{person}{Brooke~A. Simecka}.}
  \bibinfo{year}{2016}\natexlab{}.
\newblock \showarticletitle{Mental health concerns in the Digital age}.
\newblock \bibinfo{journal}{\emph{International Journal of Mental Health and
  Addiction}} \bibinfo{volume}{15}, \bibinfo{number}{3} (\bibinfo{year}{2016}),
  \bibinfo{pages}{604–613}.
\newblock
\urldef\tempurl%
\url{https://doi.org/10.1007/s11469-016-9684-0}
\showDOI{\tempurl}


\bibitem[Setlur and Tory(2022)]%
        {setlur_how_2022}
\bibfield{author}{\bibinfo{person}{Vidya Setlur} {and} \bibinfo{person}{Melanie
  Tory}.} \bibinfo{year}{2022}\natexlab{}.
\newblock \showarticletitle{How do you {Converse} with an {Analytical}
  {Chatbot}? {Revisiting} {Gricean} {Maxims} for {Designing} {Analytical}
  {Conversational} {Behavior}}. In \bibinfo{booktitle}{\emph{Proceedings of the
  2022 {CHI} {Conference} on {Human} {Factors} in {Computing} {Systems}}}
  \emph{(\bibinfo{series}{{CHI} '22})}. \bibinfo{publisher}{Association for
  Computing Machinery}, \bibinfo{address}{New York, NY, USA},
  \bibinfo{pages}{1--17}.
\newblock
\showISBNx{978-1-4503-9157-3}
\urldef\tempurl%
\url{https://doi.org/10.1145/3491102.3501972}
\showDOI{\tempurl}


\bibitem[Shani et~al\mbox{.}(2022)]%
        {shani_alexa_2022}
\bibfield{author}{\bibinfo{person}{Chen Shani}, \bibinfo{person}{Alexander
  Libov}, \bibinfo{person}{Sofia Tolmach}, \bibinfo{person}{Liane Lewin-Eytan},
  \bibinfo{person}{Yoelle Maarek}, {and} \bibinfo{person}{Dafna Shahaf}.}
  \bibinfo{year}{2022}\natexlab{}.
\newblock \showarticletitle{“{Alexa}, {Do} {You} {Want} to {Build} a
  {Snowman}?” {Characterizing} {Playful} {Requests} to {Conversational}
  {Agents}}. In \bibinfo{booktitle}{\emph{Extended {Abstracts} of the 2022
  {CHI} {Conference} on {Human} {Factors} in {Computing} {Systems}}}
  \emph{(\bibinfo{series}{{CHI} {EA} '22})}. \bibinfo{publisher}{Association
  for Computing Machinery}, \bibinfo{address}{New York, NY, USA},
  \bibinfo{pages}{1--7}.
\newblock
\showISBNx{978-1-4503-9156-6}
\urldef\tempurl%
\url{https://doi.org/10.1145/3491101.3519870}
\showDOI{\tempurl}


\bibitem[Shedroff(2000)]%
        {shedroff200011}
\bibfield{author}{\bibinfo{person}{Nathan Shedroff}.}
  \bibinfo{year}{2000}\natexlab{}.
\newblock \showarticletitle{11 Information Interaction Design: A Unified Field
  Theory of Design}.
\newblock \bibinfo{journal}{\emph{Information design}} (\bibinfo{year}{2000}),
  \bibinfo{pages}{267}.
\newblock


\bibitem[Stables(2022)]%
        {stables_2022}
\bibfield{author}{\bibinfo{person}{James Stables}.}
  \bibinfo{year}{2022}\natexlab{}.
\newblock \bibinfo{title}{Alexa brief mode explained: How to turn it on and how
  it works}.
\newblock
\newblock
\urldef\tempurl%
\url{https://www.the-ambient.com/how-to/how-to-use-alexa-brief-mode-500}
\showURL{%
\tempurl}


\bibitem[Staggers and Norcio(1993)]%
        {staggers_mental_1993}
\bibfield{author}{\bibinfo{person}{Nancy Staggers} {and} \bibinfo{person}{A.~F.
  Norcio}.} \bibinfo{year}{1993}\natexlab{}.
\newblock \showarticletitle{Mental models: concepts for human-computer
  interaction research}.
\newblock \bibinfo{journal}{\emph{International Journal of Man-Machine
  Studies}} \bibinfo{volume}{38}, \bibinfo{number}{4} (\bibinfo{date}{April}
  \bibinfo{year}{1993}), \bibinfo{pages}{587--605}.
\newblock
\showISSN{0020-7373}
\urldef\tempurl%
\url{https://doi.org/10.1006/imms.1993.1028}
\showDOI{\tempurl}


\bibitem[Striegl et~al\mbox{.}(2021)]%
        {striegl}
\bibfield{author}{\bibinfo{person}{Julian Striegl}, \bibinfo{person}{David
  Gollasch}, \bibinfo{person}{Claudia Loitsch}, {and} \bibinfo{person}{Gerhard
  Weber}.} \bibinfo{year}{2021}\natexlab{}.
\newblock \showarticletitle{Designing VUIs for Social Assistance Robots for
  People with Dementia}. In \bibinfo{booktitle}{\emph{Mensch Und Computer
  2021}} (Ingolstadt, Germany) \emph{(\bibinfo{series}{MuC '21})}.
  \bibinfo{publisher}{Association for Computing Machinery},
  \bibinfo{address}{New York, NY, USA}, \bibinfo{pages}{145–155}.
\newblock
\showISBNx{9781450386456}
\urldef\tempurl%
\url{https://doi.org/10.1145/3473856.3473887}
\showDOI{\tempurl}


\bibitem[Suhm(2003)]%
        {suhm_towards_2003}
\bibfield{author}{\bibinfo{person}{Bernhard Suhm}.}
  \bibinfo{year}{2003}\natexlab{}.
\newblock \showarticletitle{Towards best practices for speech user interface
  design}. In \bibinfo{booktitle}{\emph{8th {European} {Conference} on {Speech}
  {Communication} and {Technology} ({Eurospeech} 2003)}}.
  \bibinfo{publisher}{ISCA}, \bibinfo{pages}{2217--2220}.
\newblock
\urldef\tempurl%
\url{https://doi.org/10.21437/Eurospeech.2003-621}
\showDOI{\tempurl}


\bibitem[Sun et~al\mbox{.}(2020)]%
        {sun2020alexa}
\bibfield{author}{\bibinfo{person}{Ke Sun}, \bibinfo{person}{Chen Chen}, {and}
  \bibinfo{person}{Xinyu Zhang}.} \bibinfo{year}{2020}\natexlab{}.
\newblock \showarticletitle{" Alexa, stop spying on me!" speech privacy
  protection against voice assistants}. In
  \bibinfo{booktitle}{\emph{Proceedings of the 18th conference on embedded
  networked sensor systems}}. \bibinfo{pages}{298--311}.
\newblock


\bibitem[Tabassum et~al\mbox{.}(2019)]%
        {alwayslistening2}
\bibfield{author}{\bibinfo{person}{Madiha Tabassum}, \bibinfo{person}{Tomasz
  Kosi\'{n}ski}, \bibinfo{person}{Alisa Frik}, \bibinfo{person}{Nathan Malkin},
  \bibinfo{person}{Primal Wijesekera}, \bibinfo{person}{Serge Egelman}, {and}
  \bibinfo{person}{Heather~Richter Lipford}.} \bibinfo{year}{2019}\natexlab{}.
\newblock \showarticletitle{Investigating Users' Preferences and Expectations
  for Always-Listening Voice Assistants}.
\newblock \bibinfo{journal}{\emph{Proc. ACM Interact. Mob. Wearable Ubiquitous
  Technol.}} \bibinfo{volume}{3}, \bibinfo{number}{4}, Article
  \bibinfo{articleno}{153} (\bibinfo{date}{dec} \bibinfo{year}{2019}),
  \bibinfo{numpages}{23}~pages.
\newblock
\urldef\tempurl%
\url{https://doi.org/10.1145/3369807}
\showDOI{\tempurl}


\bibitem[Tarafdar et~al\mbox{.}(2010)]%
        {tarafdar_tu_ragu_nathan_2010}
\bibfield{author}{\bibinfo{person}{Monideepa Tarafdar}, \bibinfo{person}{Qiang
  Tu}, {and} \bibinfo{person}{T.~S. Ragu-Nathan}.}
  \bibinfo{year}{2010}\natexlab{}.
\newblock \showarticletitle{Impact of technostress on end-user satisfaction and
  performance}.
\newblock \bibinfo{journal}{\emph{Journal of Management Information Systems}}
  \bibinfo{volume}{27}, \bibinfo{number}{3} (\bibinfo{year}{2010}),
  \bibinfo{pages}{303–334}.
\newblock
\urldef\tempurl%
\url{https://doi.org/10.2753/mis0742-1222270311}
\showDOI{\tempurl}


\bibitem[Team(2017)]%
        {siri_team_2017}
\bibfield{author}{\bibinfo{person}{Siri Team}.}
  \bibinfo{year}{2017}\natexlab{}.
\newblock \bibinfo{title}{Deep learning for siri's voice: On-device deep
  mixture density networks for hybrid unit selection synthesis}.
\newblock
\newblock
\urldef\tempurl%
\url{https://machinelearning.apple.com/research/siri-voices}
\showURL{%
\tempurl}


\bibitem[Tentori et~al\mbox{.}(2009)]%
        {ambientdisplay}
\bibfield{author}{\bibinfo{person}{M. Tentori}, \bibinfo{person}{Daniela
  Segura}, {and} \bibinfo{person}{Jesus Favela}.}
  \bibinfo{year}{2009}\natexlab{}.
\newblock \showarticletitle{Monitoring hospital patients using ambient
  displays}.
\newblock \bibinfo{journal}{\emph{Mobile Health Solutions for Biomedical
  Applications}} (\bibinfo{date}{01} \bibinfo{year}{2009}),
  \bibinfo{pages}{143--158}.
\newblock
\urldef\tempurl%
\url{https://doi.org/10.4018/978-1-60566-332-6.ch008}
\showDOI{\tempurl}


\bibitem[Terzopoulos and Satratzemi(2020)]%
        {terzopoulos_satratzemi_2020}
\bibfield{author}{\bibinfo{person}{George Terzopoulos} {and}
  \bibinfo{person}{Maya Satratzemi}.} \bibinfo{year}{2020}\natexlab{}.
\newblock \showarticletitle{Voice assistants and smart speakers in Everyday
  Life and in Education}.
\newblock \bibinfo{journal}{\emph{Informatics in Education}}
  (\bibinfo{date}{Sep} \bibinfo{year}{2020}), \bibinfo{pages}{473–490}.
\newblock
\urldef\tempurl%
\url{https://doi.org/10.15388/infedu.2020.21}
\showDOI{\tempurl}


\bibitem[Trajkova and Martin-Hammond(2020)]%
        {alexatoy}
\bibfield{author}{\bibinfo{person}{Milka Trajkova} {and}
  \bibinfo{person}{Aqueasha Martin-Hammond}.} \bibinfo{year}{2020}\natexlab{}.
\newblock \showarticletitle{"Alexa is a Toy": Exploring Older Adults' Reasons
  for Using, Limiting, and Abandoning Echo}. In
  \bibinfo{booktitle}{\emph{Proceedings of the 2020 CHI Conference on Human
  Factors in Computing Systems}} (Honolulu, HI, USA)
  \emph{(\bibinfo{series}{CHI '20})}. \bibinfo{publisher}{Association for
  Computing Machinery}, \bibinfo{address}{New York, NY, USA},
  \bibinfo{pages}{1–13}.
\newblock
\showISBNx{9781450367080}
\urldef\tempurl%
\url{https://doi.org/10.1145/3313831.3376760}
\showDOI{\tempurl}


\bibitem[Ulijn et~al\mbox{.}(2012)]%
        {ulijn_strother_fazal_2012}
\bibfield{author}{\bibinfo{person}{Jan~M. Ulijn}, \bibinfo{person}{Judith~B.
  Strother}, {and} \bibinfo{person}{Zohra Fazal}.}
  \bibinfo{year}{2012}\natexlab{}.
\newblock \bibinfo{booktitle}{\emph{Chapter 2}}.
\newblock \bibinfo{publisher}{IEEE}.
\newblock


\bibitem[Vacher et~al\mbox{.}(2015)]%
        {vacher_evaluation_2015}
\bibfield{author}{\bibinfo{person}{Michel Vacher}, \bibinfo{person}{Sybille
  Caffiau}, \bibinfo{person}{François Portet}, \bibinfo{person}{Brigitte
  Meillon}, \bibinfo{person}{Camille Roux}, \bibinfo{person}{Elena Elias},
  \bibinfo{person}{Benjamin Lecouteux}, {and} \bibinfo{person}{Pedro
  Chahuara}.} \bibinfo{year}{2015}\natexlab{}.
\newblock \showarticletitle{Evaluation of a {Context}-{Aware} {Voice}
  {Interface} for {Ambient} {Assisted} {Living}: {Qualitative} {User} {Study}
  vs. {Quantitative} {System} {Evaluation}}.
\newblock \bibinfo{journal}{\emph{ACM Transactions on Accessible Computing}}
  \bibinfo{volume}{7}, \bibinfo{number}{2} (\bibinfo{date}{July}
  \bibinfo{year}{2015}), \bibinfo{pages}{1--36}.
\newblock
\showISSN{1936-7228, 1936-7236}
\urldef\tempurl%
\url{https://doi.org/10.1145/2738047}
\showDOI{\tempurl}


\bibitem[Voit et~al\mbox{.}(2020)]%
        {voit}
\bibfield{author}{\bibinfo{person}{Alexandra Voit}, \bibinfo{person}{Jasmin
  Niess}, \bibinfo{person}{Caroline Eckerth}, \bibinfo{person}{Maike Ernst},
  \bibinfo{person}{Henrike Weing\"{a}rtner}, {and} \bibinfo{person}{Pawe\l{}~W.
  Wo\'{z}niak}.} \bibinfo{year}{2020}\natexlab{}.
\newblock \showarticletitle{‘It’s Not a Romantic Relationship’: Stories
  of Adoption and Abandonment of Smart Speakers at Home}. In
  \bibinfo{booktitle}{\emph{19th International Conference on Mobile and
  Ubiquitous Multimedia}} (Essen, Germany) \emph{(\bibinfo{series}{MUM '20})}.
  \bibinfo{publisher}{Association for Computing Machinery},
  \bibinfo{address}{New York, NY, USA}, \bibinfo{pages}{71–82}.
\newblock
\showISBNx{9781450388702}
\urldef\tempurl%
\url{https://doi.org/10.1145/3428361.3428469}
\showDOI{\tempurl}


\bibitem[Wang et~al\mbox{.}(2020)]%
        {wang_alexa_2020}
\bibfield{author}{\bibinfo{person}{Jinping Wang}, \bibinfo{person}{Hyun Yang},
  \bibinfo{person}{Ruosi Shao}, \bibinfo{person}{Saeed Abdullah}, {and}
  \bibinfo{person}{S.~Shyam Sundar}.} \bibinfo{year}{2020}\natexlab{}.
\newblock \showarticletitle{Alexa as {Coach}: {Leveraging} {Smart} {Speakers}
  to {Build} {Social} {Agents} that {Reduce} {Public} {Speaking} {Anxiety}}. In
  \bibinfo{booktitle}{\emph{Proceedings of the 2020 {CHI} {Conference} on
  {Human} {Factors} in {Computing} {Systems}}} \emph{(\bibinfo{series}{{CHI}
  '20})}. \bibinfo{publisher}{Association for Computing Machinery},
  \bibinfo{address}{New York, NY, USA}, \bibinfo{pages}{1--13}.
\newblock
\showISBNx{978-1-4503-6708-0}
\urldef\tempurl%
\url{https://doi.org/10.1145/3313831.3376561}
\showDOI{\tempurl}


\bibitem[Wei and Landay(2018)]%
        {wei_evaluating_2018}
\bibfield{author}{\bibinfo{person}{Zhuxiaona Wei} {and}
  \bibinfo{person}{James~A. Landay}.} \bibinfo{year}{2018}\natexlab{}.
\newblock \showarticletitle{Evaluating {Speech}-{Based} {Smart} {Devices}
  {Using} {New} {Usability} {Heuristics}}.
\newblock \bibinfo{journal}{\emph{IEEE Pervasive Computing}}
  \bibinfo{volume}{17}, \bibinfo{number}{2} (\bibinfo{date}{April}
  \bibinfo{year}{2018}), \bibinfo{pages}{84--96}.
\newblock
\showISSN{1536-1268, 1558-2590}
\urldef\tempurl%
\url{https://doi.org/10.1109/MPRV.2018.022511249}
\showDOI{\tempurl}


\bibitem[Weiser(1991)]%
        {weiser_computer_nodate}
\bibfield{author}{\bibinfo{person}{Mark Weiser}.}
  \bibinfo{year}{1991}\natexlab{}.
\newblock \showarticletitle{The {Computer} for the 21st {Century}}.
\newblock  (\bibinfo{year}{1991}), \bibinfo{pages}{8}.
\newblock


\bibitem[Weiser(1998)]%
        {weiserschool}
\bibfield{author}{\bibinfo{person}{Mark Weiser}.}
  \bibinfo{year}{1998}\natexlab{}.
\newblock \showarticletitle{The Future of Ubiquitous Computing on Campus}.
\newblock \bibinfo{journal}{\emph{Commun. ACM}} \bibinfo{volume}{41},
  \bibinfo{number}{1} (\bibinfo{date}{jan} \bibinfo{year}{1998}),
  \bibinfo{pages}{41–42}.
\newblock
\showISSN{0001-0782}
\urldef\tempurl%
\url{https://doi.org/10.1145/268092.268108}
\showDOI{\tempurl}


\bibitem[Weiser and Brown(1995)]%
        {weiser_designing_1995}
\bibfield{author}{\bibinfo{person}{Mark Weiser} {and}
  \bibinfo{person}{John~Seely Brown}.} \bibinfo{year}{1995}\natexlab{}.
\newblock \bibinfo{title}{Designing {Calm} {Technology}}.
\newblock
\newblock
\urldef\tempurl%
\url{https://calmtech.com/papers/designing-calm-technology.html}
\showURL{%
\tempurl}


\bibitem[Weiser and Brown(1996)]%
        {weiser_coming_1996}
\bibfield{author}{\bibinfo{person}{Mark Weiser} {and}
  \bibinfo{person}{John~Seely Brown}.} \bibinfo{year}{1996}\natexlab{}.
\newblock \bibinfo{title}{The {Coming} {Age} of {Calm} {Technology}}.
\newblock
\newblock
\urldef\tempurl%
\url{https://calmtech.com/papers/coming-age-calm-technology.html}
\showURL{%
\tempurl}


\bibitem[Yang et~al\mbox{.}(2019)]%
        {yang_understanding_2019}
\bibfield{author}{\bibinfo{person}{Xi Yang}, \bibinfo{person}{Marco
  Aurisicchio}, {and} \bibinfo{person}{Weston Baxter}.}
  \bibinfo{year}{2019}\natexlab{}.
\newblock \showarticletitle{Understanding {Affective} {Experiences} with
  {Conversational} {Agents}}. In \bibinfo{booktitle}{\emph{Proceedings of the
  2019 {CHI} {Conference} on {Human} {Factors} in {Computing} {Systems}}}
  \emph{(\bibinfo{series}{{CHI} '19})}. \bibinfo{publisher}{Association for
  Computing Machinery}, \bibinfo{address}{New York, NY, USA},
  \bibinfo{pages}{1--12}.
\newblock
\showISBNx{978-1-4503-5970-2}
\urldef\tempurl%
\url{https://doi.org/10.1145/3290605.3300772}
\showDOI{\tempurl}


\bibitem[Yu et~al\mbox{.}(2018)]%
        {yu_shane_schlosser_o’brien_allen_abramson_flynn_2018}
\bibfield{author}{\bibinfo{person}{Christina Yu}, \bibinfo{person}{Howard
  Shane}, \bibinfo{person}{Ralf~W. Schlosser}, \bibinfo{person}{Amanda
  O’Brien}, \bibinfo{person}{Anna Allen}, \bibinfo{person}{Jennifer
  Abramson}, {and} \bibinfo{person}{Suzanne Flynn}.}
  \bibinfo{year}{2018}\natexlab{}.
\newblock \showarticletitle{An exploratory study of speech-language
  pathologists using the Echo Show™ to deliver visual supports}.
\newblock \bibinfo{journal}{\emph{Advances in Neurodevelopmental Disorders}}
  \bibinfo{volume}{2}, \bibinfo{number}{3} (\bibinfo{year}{2018}),
  \bibinfo{pages}{286–292}.
\newblock
\urldef\tempurl%
\url{https://doi.org/10.1007/s41252-018-0075-3}
\showDOI{\tempurl}


\bibitem[Zhang et~al\mbox{.}(2022a)]%
        {zhang_aware_2022}
\bibfield{author}{\bibinfo{person}{Xinlei Zhang}, \bibinfo{person}{Zixiong Su},
  {and} \bibinfo{person}{Jun Rekimoto}.} \bibinfo{year}{2022}\natexlab{a}.
\newblock \showarticletitle{Aware: {Intuitive} {Device} {Activation} {Using}
  {Prosody} for {Natural} {Voice} {Interactions}}. In
  \bibinfo{booktitle}{\emph{Proceedings of the 2022 {CHI} {Conference} on
  {Human} {Factors} in {Computing} {Systems}}} \emph{(\bibinfo{series}{{CHI}
  '22})}. \bibinfo{publisher}{Association for Computing Machinery},
  \bibinfo{address}{New York, NY, USA}, \bibinfo{pages}{1--16}.
\newblock
\showISBNx{978-1-4503-9157-3}
\urldef\tempurl%
\url{https://doi.org/10.1145/3491102.3517687}
\showDOI{\tempurl}


\bibitem[Zhang et~al\mbox{.}(2022b)]%
        {zhang_storybuddy_2022}
\bibfield{author}{\bibinfo{person}{Zheng Zhang}, \bibinfo{person}{Ying Xu},
  \bibinfo{person}{Yanhao Wang}, \bibinfo{person}{Bingsheng Yao},
  \bibinfo{person}{Daniel Ritchie}, \bibinfo{person}{Tongshuang Wu},
  \bibinfo{person}{Mo Yu}, \bibinfo{person}{Dakuo Wang}, {and}
  \bibinfo{person}{Toby Jia-Jun Li}.} \bibinfo{year}{2022}\natexlab{b}.
\newblock \showarticletitle{{StoryBuddy}: {A} {Human}-{AI} {Collaborative}
  {Chatbot} for {Parent}-{Child} {Interactive} {Storytelling} with {Flexible}
  {Parental} {Involvement}}. In \bibinfo{booktitle}{\emph{Proceedings of the
  2022 {CHI} {Conference} on {Human} {Factors} in {Computing} {Systems}}}
  \emph{(\bibinfo{series}{{CHI} '22})}. \bibinfo{publisher}{Association for
  Computing Machinery}, \bibinfo{address}{New York, NY, USA},
  \bibinfo{pages}{1--21}.
\newblock
\showISBNx{978-1-4503-9157-3}
\urldef\tempurl%
\url{https://doi.org/10.1145/3491102.3517479}
\showDOI{\tempurl}


\bibitem[Zubatiy et~al\mbox{.}(2021)]%
        {zubatiy_empowering_2021}
\bibfield{author}{\bibinfo{person}{Tamara Zubatiy}, \bibinfo{person}{Kayci~L
  Vickers}, \bibinfo{person}{Niharika Mathur}, {and}
  \bibinfo{person}{Elizabeth~D Mynatt}.} \bibinfo{year}{2021}\natexlab{}.
\newblock \showarticletitle{Empowering {Dyads} of {Older} {Adults} {With}
  {Mild} {Cognitive} {Impairment} {And} {Their} {Care} {Partners} {Using}
  {Conversational} {Agents}}. In \bibinfo{booktitle}{\emph{Proceedings of the
  2021 {CHI} {Conference} on {Human} {Factors} in {Computing} {Systems}}}
  \emph{(\bibinfo{series}{{CHI} '21})}. \bibinfo{publisher}{Association for
  Computing Machinery}, \bibinfo{address}{New York, NY, USA},
  \bibinfo{pages}{1--15}.
\newblock
\showISBNx{978-1-4503-8096-6}
\urldef\tempurl%
\url{https://doi.org/10.1145/3411764.3445124}
\showDOI{\tempurl}


\bibitem[Zwakman et~al\mbox{.}(2021)]%
        {zwakman_usability_2021}
\bibfield{author}{\bibinfo{person}{Dilawar~Shah Zwakman},
  \bibinfo{person}{Debajyoti Pal}, {and} \bibinfo{person}{Chonlameth
  Arpnikanondt}.} \bibinfo{year}{2021}\natexlab{}.
\newblock \showarticletitle{Usability {Evaluation} of {Artificial}
  {Intelligence}-{Based} {Voice} {Assistants}: {The} {Case} of {Amazon}
  {Alexa}}.
\newblock \bibinfo{journal}{\emph{Sn Computer Science}} \bibinfo{volume}{2},
  \bibinfo{number}{1} (\bibinfo{year}{2021}), \bibinfo{pages}{28}.
\newblock
\showISSN{2662-995X}
\urldef\tempurl%
\url{https://doi.org/10.1007/s42979-020-00424-4}
\showDOI{\tempurl}


\end{thebibliography}
